\documentclass[a4paper,fleqn,usenatbib]{mnras}

\usepackage[T1]{fontenc}
\usepackage{ae,aecompl}
\usepackage{graphicx}	% Including figure files
\usepackage{amsmath}	% Advanced maths commands
\usepackage{amssymb}	% Extra maths symbols

\title[Optical phase curves]{Optical phase curves as diagnostics for aerosol composition in exoplanetary atmospheres}

\author[Oreshenko, Heng \& Demory]{
Maria Oreshenko,$^{1,2}$\thanks{E-mail: maria.oreshenko@csh.unibe.ch (MO)}
Kevin Heng,$^{1}$\thanks{Email: kevin.heng@csh.unibe.ch (KH)}
Brice-Olivier Demory$^{3}$\thanks{Email: bod21@cam.ac.uk (BOD)}
\\
$^{1}$University of Bern, Center for Space and Habitability, Sidlerstrasse 5, CH-3012, Bern, Switzerland\\
$^{2}$ETH Z\"{u}rich, Institute for Astronomy, Wolfgang-Pauli-Strasse 27, CH-8093, Z\"{u}rich, Switzerland\\
$^{3}$University of Cambridge, Cavendish Laboratory, 19 JJ Thomson Avenue, Cambridge CB3 0HE, U.K.}

\date{Accepted 14th Jan 2016. Received 12th Jan 2016; in original form 9th Oct 2015.}

\pubyear{2015}

\begin{document}
\label{firstpage}
\pagerange{\pageref{firstpage}--\pageref{lastpage}}
\maketitle

% Abstract of the paper
\begin{abstract}
Optical phase curves have become one of the common probes of exoplanetary atmospheres, but the information they encode has not been fully elucidated.  Building on a diverse body of work, we upgrade the Flexible Modeling System (FMS) to include scattering in the two-stream, dual-band approximation and generate plausible, three-dimensional structures of irradiated atmospheres to study the radiative effects of aerosols or condensates.  In the optical, we treat the scattering of starlight using a generalisation of Beer's law that allows for a finite Bond albedo to be prescribed.  In the infrared, we implement the two-stream solutions and include scattering via an infrared scattering parameter.  We present a suite of four-parameter general circulation models for Kepler-7b and demonstrate that its climatology is expected to be robust to variations in optical and infrared scattering.  The westward and eastward shifts of the optical and infrared phase curves, respectively, are shown to be robust outcomes of the simulations.  Assuming micron-sized particles and a simplified treatment of local brightness, we further show that the peak offset of the optical phase curve is sensitive to the composition of the aerosols or condensates.  However, to within the measurement uncertainties, we cannot distinguish between aerosols made of silicates (enstatite or forsterite), iron, corundum or titanium oxide, based on a comparison to the measured peak offset ($41^\circ \pm 12^\circ$) of the optical phase curve of Kepler-7b.  Measuring high-precision optical phase curves will provide important constraints on the atmospheres of cloudy exoplanets and reduce degeneracies in interpreting their infrared spectra.
\end{abstract}

\begin{keywords}
planets and satellites: atmospheres -- hydrodynamics -- radiative transfer -- scattering -- methods: numerical
\end{keywords}

\section{Introduction}
\label{sect:intro}

Optical phase curves of exoplanets serve as a common probe of their atmospheres and have been recorded using the CoRoT \citep{snellen09} and Kepler Space Telescopes \citep{borucki09,welsh10,barclay12,demory13,faigler13,quintana13,shporer14,anger15,esteves15}.  If the exoplanetary atmosphere is cool enough such that the optical phase curve derives predominantly from reflected starlight, then they encode information about the properties of aerosols or condensates \citep{hd13,par13,hu15,sh15}.  Such information is complementary to what we may learn from analyzing infrared phase curves and spectra (see \citealt{crossfield15} and \citealt{hs15} for reviews).

At a basic level, the formation and existence of aerosols or condensates in an atmosphere results from a complex interplay between atmospheric dynamics, chemistry and radiation.  To solve this computational problem rigorously requires that one simulates the three-dimensional background state of temperature, velocity and mass density, and iterates it with the radiative heating from the star \citep{showman09,lewis10,kataria13}.  The thermal state of the atmosphere is in turn dependent upon its opacity function, which is determined by its constituent chemistry.  Non-equilibrium chemistry may be driven by atmospheric dynamics \citep{cs06,agundez12}.  Additionally, one has to worry about both gas- and solid-phase chemistry \citep{sh90,bs99}.  A formation theory of aerosols or condensates is set against this complex backdrop of physics and chemistry.

Unsurprisingly, all of the existing studies of aerosols or condensates in irradiated exoplanetary atmospheres employ some form of approximation or simplification.  \cite{hhps12} constructed one-dimensional temperature-pressure profiles of irradiated exoplanetary atmospheres and included the effects of scattering by aerosols.  \cite{heng12} used the model of \cite{hhps12} to study the effects of aerosols or condensates on Ohmic dissipation in hot exoplanetary atmospheres.  \cite{par13} constructed three-dimensional general circulation models (GCMs) of hot Jupiters with ``tracers", which are computational ``beads" inserted into the flow to allow one to record its local properties.  They modelled the dynamical coupling between the atmospheric flow and aerosols by assigning a particle radius, terminal velocity and Knudsen number to each tracer, but did not model the radiative forcing (absorption and scattering) of the aerosols on the thermal structure of the atmosphere.  \cite{par13} also did not consider the scattering of radiation.  \cite{hd13} ignored the three-dimensional fluid dynamical structure of the flow and instead focused on the radiative forcing of the aerosols via a prescribed albedo.  \cite{lee15} post-processed three-dimensional, but non-global, GCMs with kinetic models of cloud formation to predict distributions in the composition, sizes and number densities of aerosols, but did not calculate gas and aerosol chemistry self-consistently.  \cite{par13}, \cite{hd13} and \cite{lee15} demonstrated that micron-sized particles should be ubiquitous---or at least easily lofted---in hot Jovian atmospheres.  However, \cite{par13}, \cite{hd13} and \cite{lee15} did not explicitly model the optical phase curves.  \cite{hu15} developed a semi-analytical model that allowed them to simultaneously fit infrared and optical phase curves and obtain constraints on the Bond albedo, heat redistribution efficiency, greenhouse warming and condensation temperature of the aerosol.  \cite{gi15} solved the radiative transfer equation with multiple scattering and a detailed treatment of the optical properties of the aerosols/condensates to generate a large grid of 6-parameter optical phase curves, which they then fitted to the measured optical phase curve of Kepler-7b \citep{demory13} and concluded that small (sub-micron-sized) particles are present in its atmosphere.

Given this rich and diverse body of work, it is reasonable to contribute to the study of aerosols and condensates in irradiated exoplanetary atmospheres from a different perspective: to study the problem using a simplified GCM and compute both infrared and optical phase curves contemporaneously.  This is the goal of the present study, where we build upon the work of \cite{hmp11} and \cite{hfp11} by adding a simplified treatment of scattering to the Flexible Modeling System (FMS) GCM.  Our main conclusion is that, while the climatology of our model hot Jupiters appear to be robust to variations in optical and infrared scattering, the optical phase curves are sensitive to the chemical composition of the aerosols.  Tentatively, we conclude that the peak offsets of optical phase curves may be used as a diagnostic to constrain aerosol chemistry and composition.  In the future, this conclusion should be checked by more sophisticated calculations that include multiple scattering.

In Section \ref{sect:methods}, we describe our methods, including the equations used and how we implemented them into our upgraded GCM.  In Section \ref{sect:results}, we present a suite of GCMs customised to the hot Jupiter Kepler-7b.  In Section \ref{sect:discussion}, we discuss the implications of our results and describe opportunities for future work.

\section{Methodology}
\label{sect:methods}

\begin{figure}
\vspace{-0.2in}
\includegraphics[width=\columnwidth]{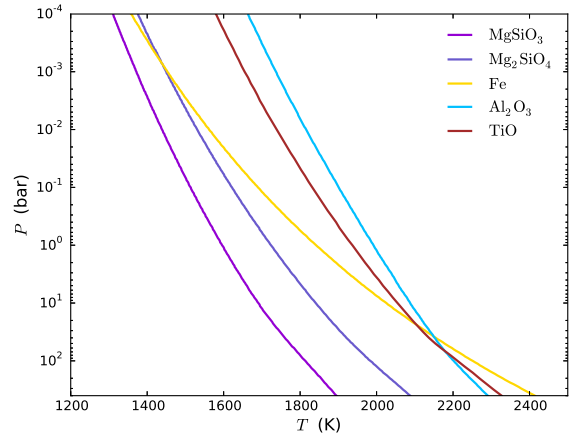}
\caption{Condensation curves of the aerosols considered in the present study.}
\vspace{-0.1in}
\label{fig:condense}
\end{figure}

\subsection{The Flexible Modelling System (FMS)}

We implement and adapt the FMS GCM of the Geophysical Fluid Dynamics Laboratory (GFDL) of Princeton University.  It solves the primitive equations of meteorology, which are essentially re-statements of the conservation of mass, momentum and energy for fluid dynamics; see, for example, \cite{vallis}.  In \cite{hmp11}, we implemented the FMS with Newtonian relaxation, which substitutes for radiative transfer by prescribing what astrophysicists commonly call a ``cooling function".  The atmosphere is then made to relax to this equilibrium state via a prescribed radiative relaxation timescale, which generally depends on temperature and pressure.  In \cite{hfp11}, we removed Newtonian relaxation from the calculation and instead implemented dual-band radiative transfer, which makes the simplifying assumption that starlight and thermal emission from the exoplanetary atmosphere are well separated in wavelength, often termed the ``shortwave" and ``longwave", respectively.  For exoplanets orbiting Sun-like stars, the shortwave and longwave are in the optical and infrared, respectively; for M stars, they may both be in the infrared.  Using the numbers listed in \cite{esteves15} for Kepler-7 (the star) and Kepler-7b (the exoplanet), we estimate, using Wien's law, that the star emits at a peak wavelength of about 0.5 $\mu$m, while the exoplanet emits at a peak wavelength of about 2 $\mu$m.

We will not repeat the technical details of our implementation of the FMS here, except when they are necessary to demonstrate a specific point, and rather refer the reader to \cite{hmp11} and \cite{hfp11}.  In \cite{php12}, we also used the computational setup in \cite{hfp11} to study a suite of tidally-locked gas-giant exoplanets with different insolations and the absence/presence of a temperature inversion (enforced via an enhanced shortwave opacity).  The FMS has also been used to study exoplanetary atmospheres under different circumstances.  \cite{pierrehumbert11} postulated the existence of an ``eyeball Earth"---a mostly frozen, water-ice exoplanet with a stable pool of water at its substellar point---using FMS GCMs.  \cite{ms10} and \cite{hv11} explored variations on a theme of tidally-locked, Earth-like exoplanets.  \cite{ka15} used the FMS to study thermal phase curves of dry, tidally-locked terrestrial exoplanets.

In the present study, our goal is to upgrade the setup in \cite{hfp11} to include scattering both in the shortwave/optical and longwave/infrared.  For the shortwave, we implement a generalisation of Beer's law that includes the scattering of starlight, to be prescribed via the Bond albedo.  For the longwave, we implement analytical two-stream solutions that allow for the scattering of infrared thermal emission via a ``scattering parameter".  The mathematical formalism and equations used to implement these upgrades have previously been described in \cite{hml14}, but we will summarise and review them, for convenience, in the following two sub-sections.

\subsection{Generalised Beer's law (optical scattering)}

Traditionally, Beer's law describes the exponential diminution of the flux of starlight as it penetrates a purely-absorbing atmosphere, where the exponent is the optical depth (multiplied by a dimensionless coefficient).  \cite{hml14} generalised Beer's law to include non-isotropic scattering in the shortwave,
\begin{equation}
F_{\rm S} = F_{\rm TOA} \left( 1 - A_{\rm B} \right) \exp{\left[ \frac{\kappa_{\rm S} m}{\mu (n+1)\beta_{\rm S}} \right]},
\label{eq:beers}
\end{equation}
where $F_{\rm TOA}$ is the stellar constant (zero albedo), $A_{\rm B}$ is the Bond albedo and $\mu$ is the cosine of the zenith angle (which is defined to be negative for incoming radiation).  The shortwave opacity is parameterized to be 
\begin{equation}
\kappa_{\rm S}=\kappa_{\rm S_0} \left( \frac{\tilde{m}}{\tilde{m}_0} \right)^n,
\end{equation}
where $\kappa_{\rm S_0}$ is a normalisation factor, $\tilde{m}$ is the column mass, $\tilde{m}_0$ is the column mass at the bottom of the simulation domain and $n$ is a dimensionless index.  Furthermore, the incident starlight or shortwave flux ($F_{\rm S}$) is geometrically diluted across latitude and longitude by cosine functions; see equation (23) of \cite{hfp11}.

\cite{hml14} previously showed that the shortwave scattering parameter ($\beta_{\rm S}$) and the Bond albedo are related via 
\begin{equation}
\beta_{\rm S}= \frac{1-A_{\rm B}}{1+A_{\rm B}}.
\end{equation}

For the suite of simulations presented in this study, we use $\mu=-1$ (to represent the radial rays of the two-stream approximation), $\kappa_{\rm S_0} = 0.005$ cm$^2$ g$^{-1}$, $n=0$ and $P_0 = 1$ kbar.  For simplicity and for comparison with \cite{php12}, we have assumed a constant shortwave opacity across pressure ($n=0$).  The bottom simulation domain of $P_0=1$ kbar is chosen such that it resides deeply enough to not artificially interfere with the photospheric region.  We note that the column mass and pressure are related via $P = \tilde{m} g$ and $P_0 = \tilde{m}_0 g$, where $g$ is the surface gravity.  We explore variations in the extinction of starlight across depth by varying the value of the Bond albedo: $A_{\rm B}=0, 0.1$ and 0.5.  Our $A_{\rm B}=0$ runs may be interpreted both as the pure absorption limit, which was previously explored in \cite{hfp11} and \cite{php12}, and also as a model in which the Bond albedo is undetectably small.  \cite{hd13} have previously showed that the geometric albedos of hot Jupiters are typically $\lesssim 0.1$ (see also \citealt{sc15}), while Kepler-7b has a geometric albedo of $0.35 \pm 0.02$ \citep{demory13}.

\subsection{Two-stream equations with scattering (infrared scattering)}

\begin{figure*}
\minipage{0.33\textwidth}
\includegraphics[width=\textwidth]{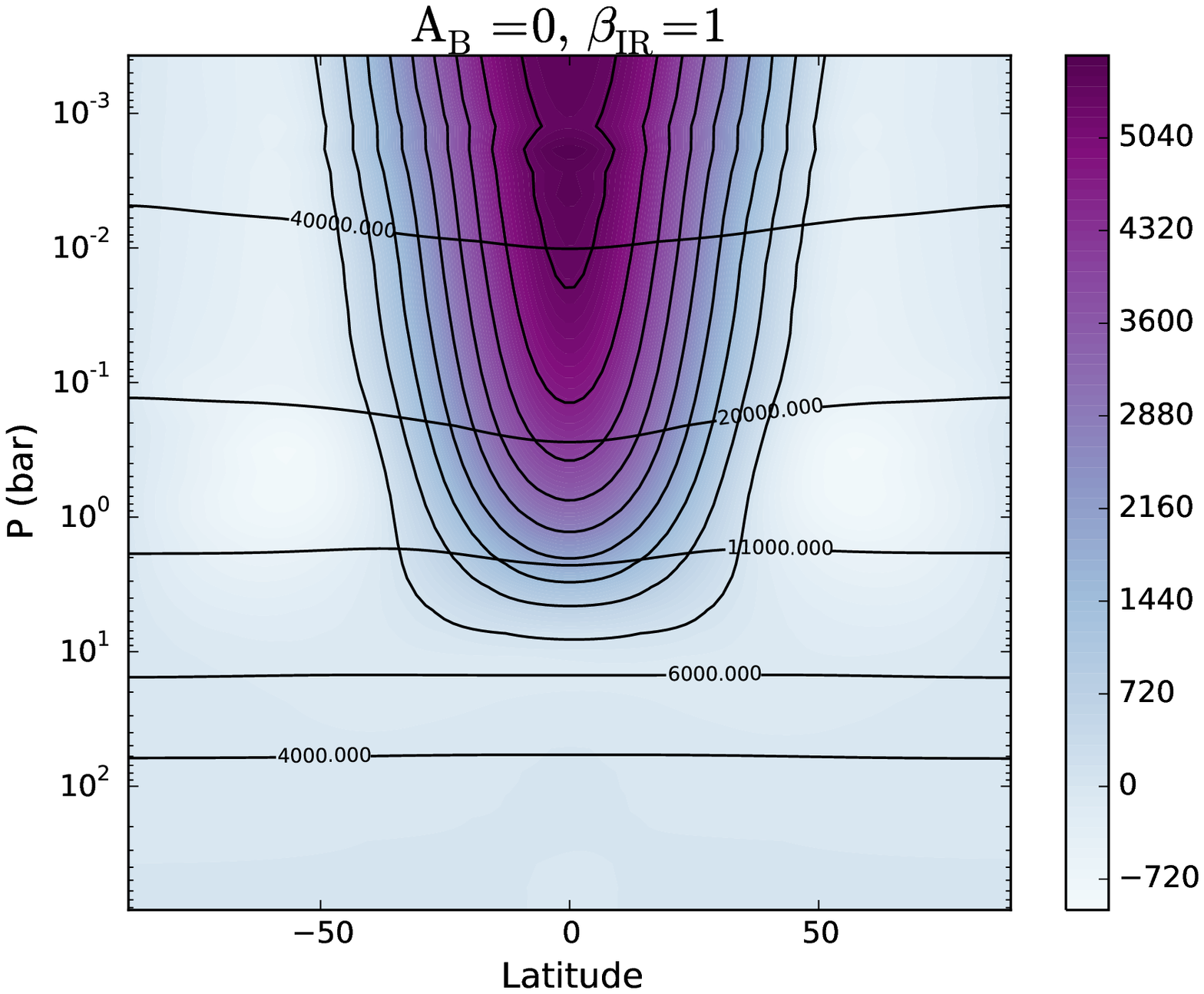}
\endminipage\hfill
\minipage{0.33\textwidth}
\includegraphics[width=\textwidth]{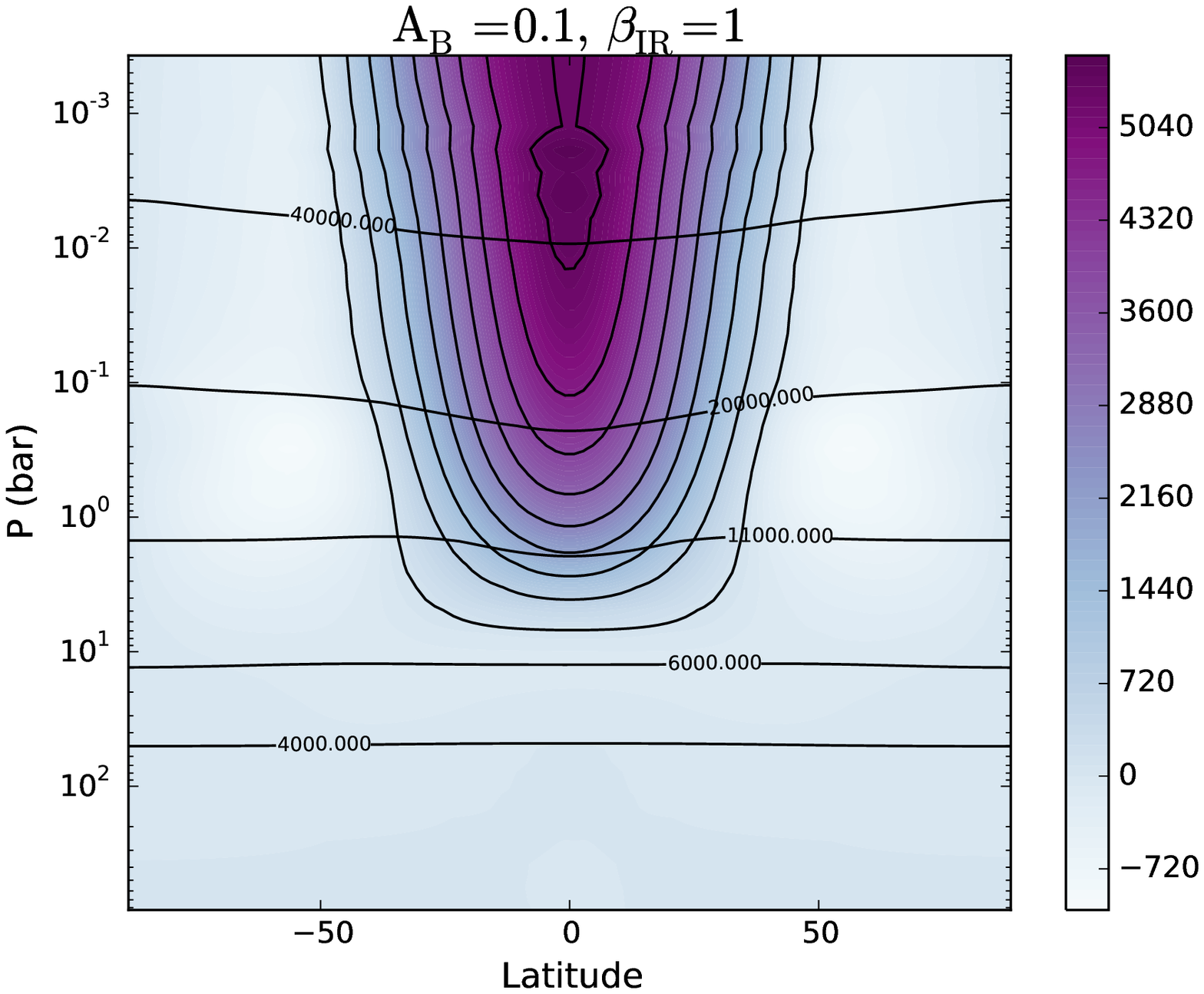}
\endminipage\hfill
\minipage{0.33\textwidth}
\includegraphics[width=\textwidth]{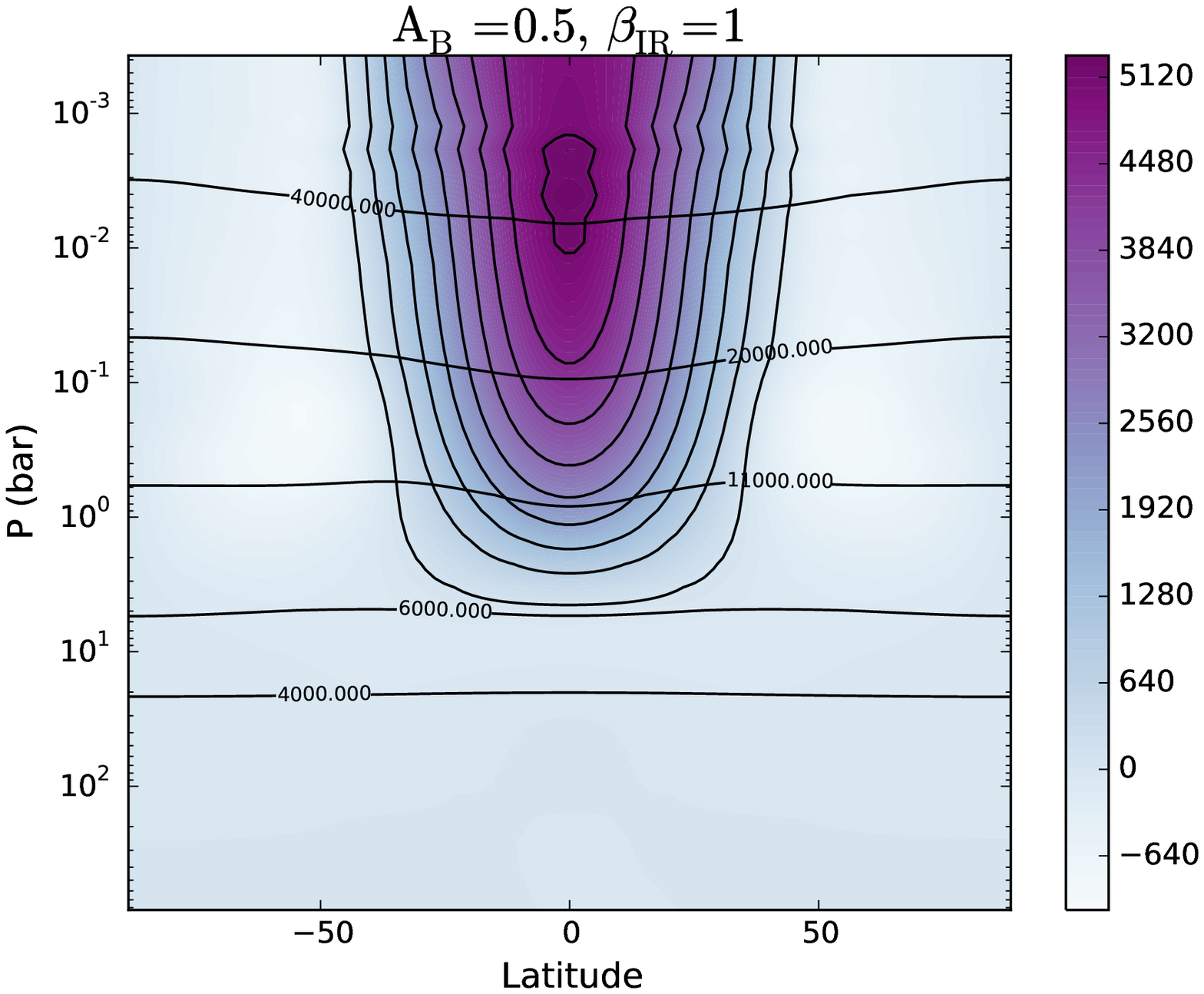}
\endminipage\hfill
\minipage{0.33\textwidth}
\includegraphics[width=\textwidth]{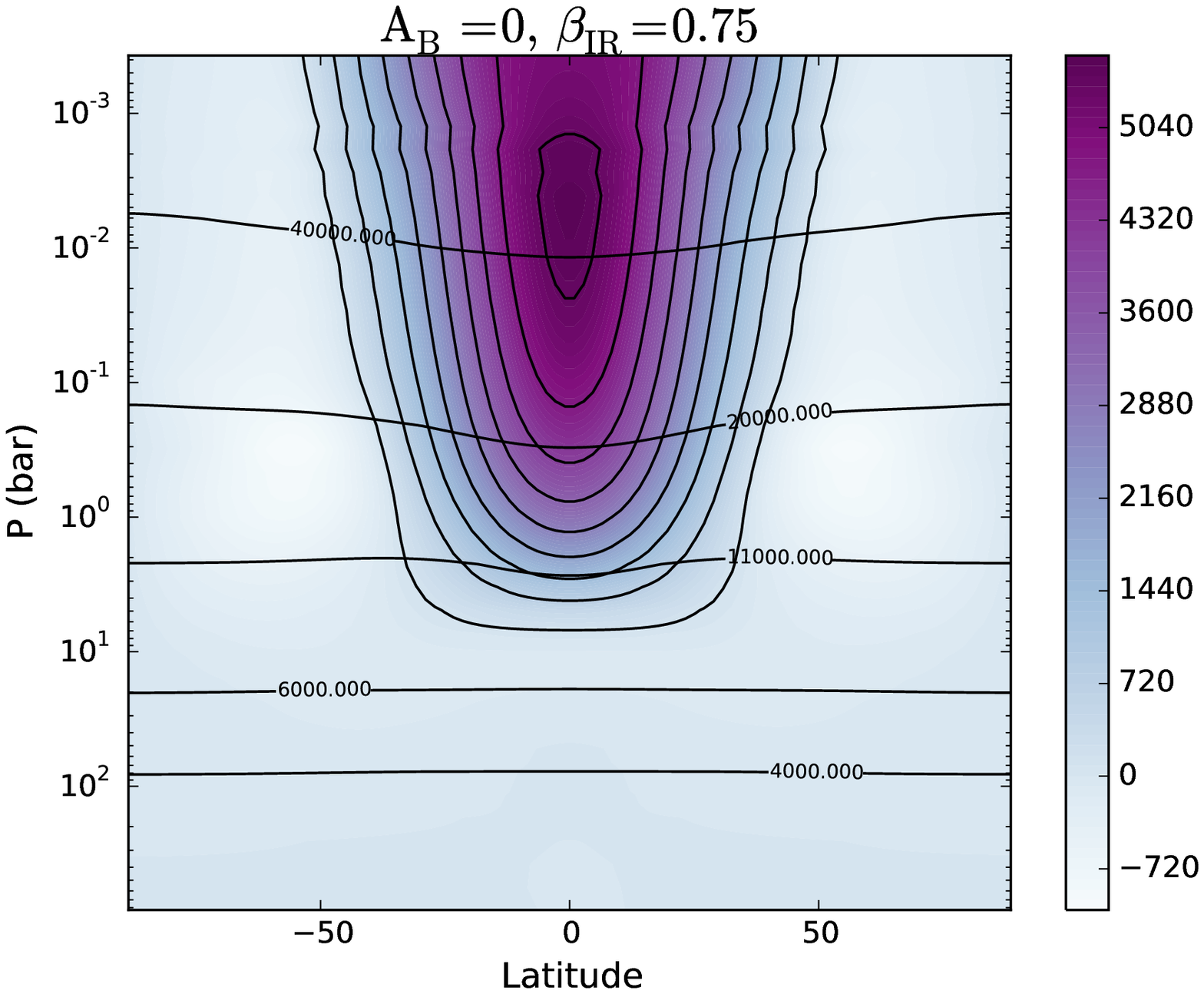}
\endminipage\hfill
\minipage{0.33\textwidth}
\includegraphics[width=\textwidth]{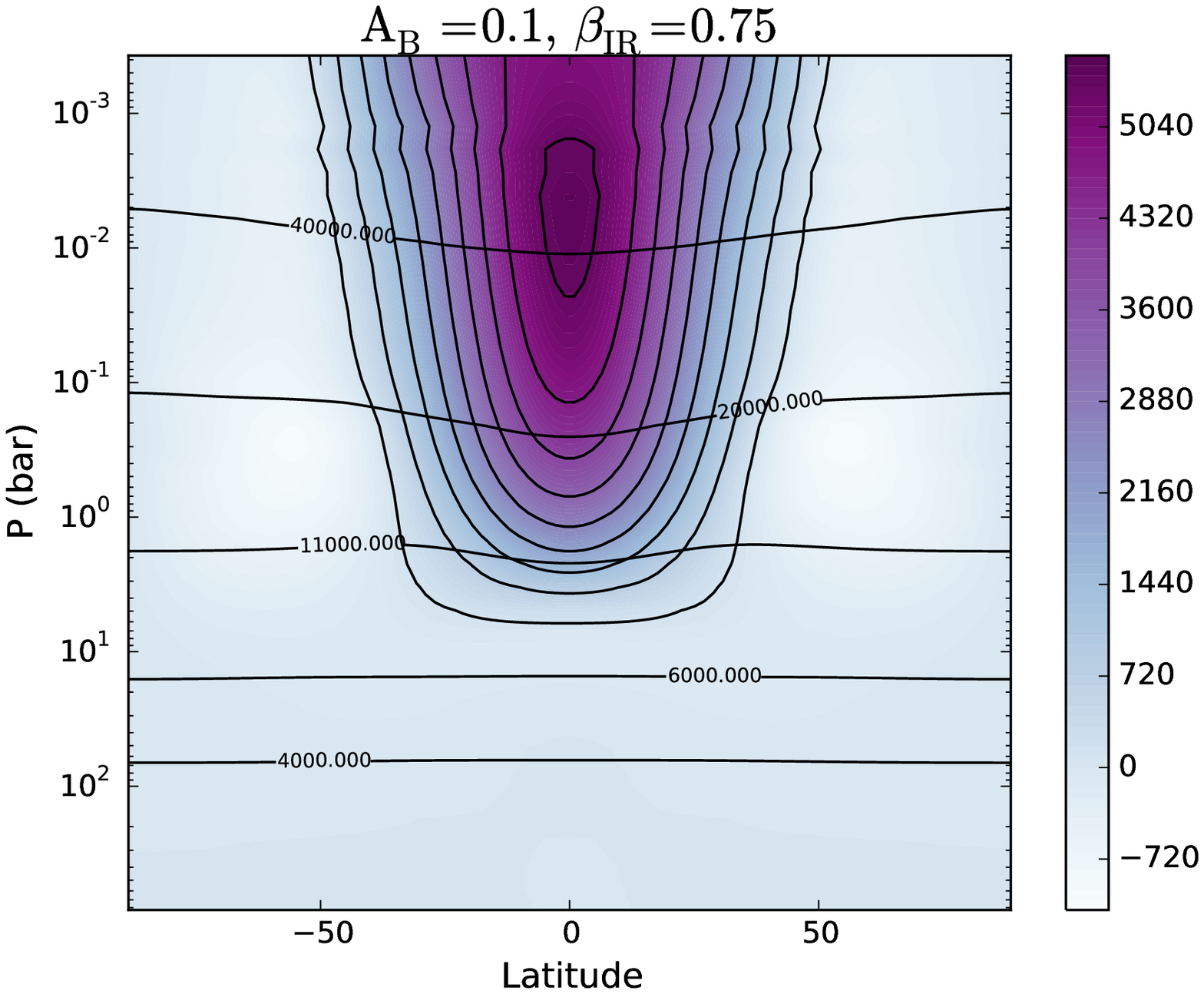}
\endminipage\hfill
\minipage{0.33\textwidth}
\includegraphics[width=\textwidth]{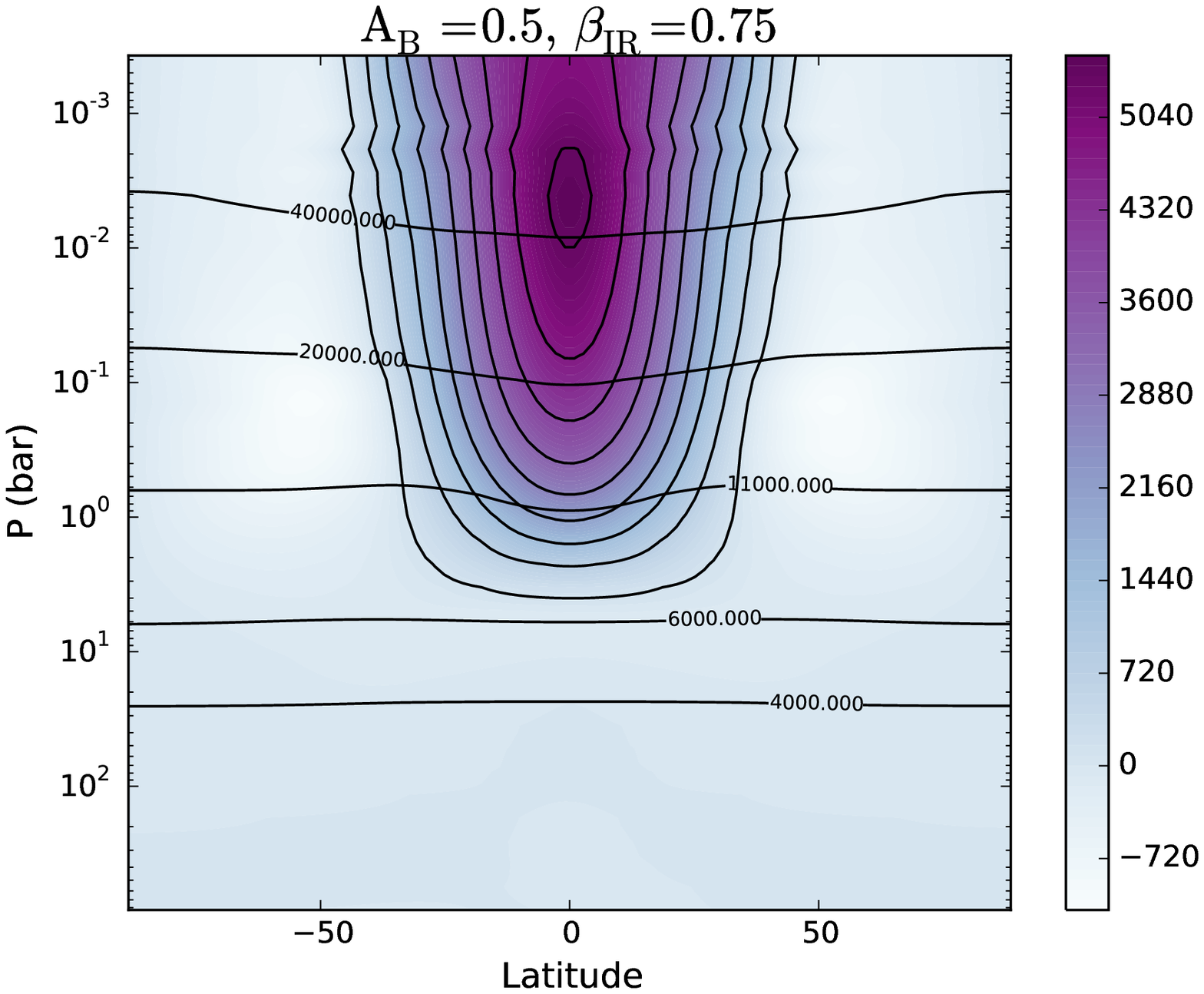}
\endminipage\hfill
\minipage{0.33\textwidth}
\includegraphics[width=\textwidth]{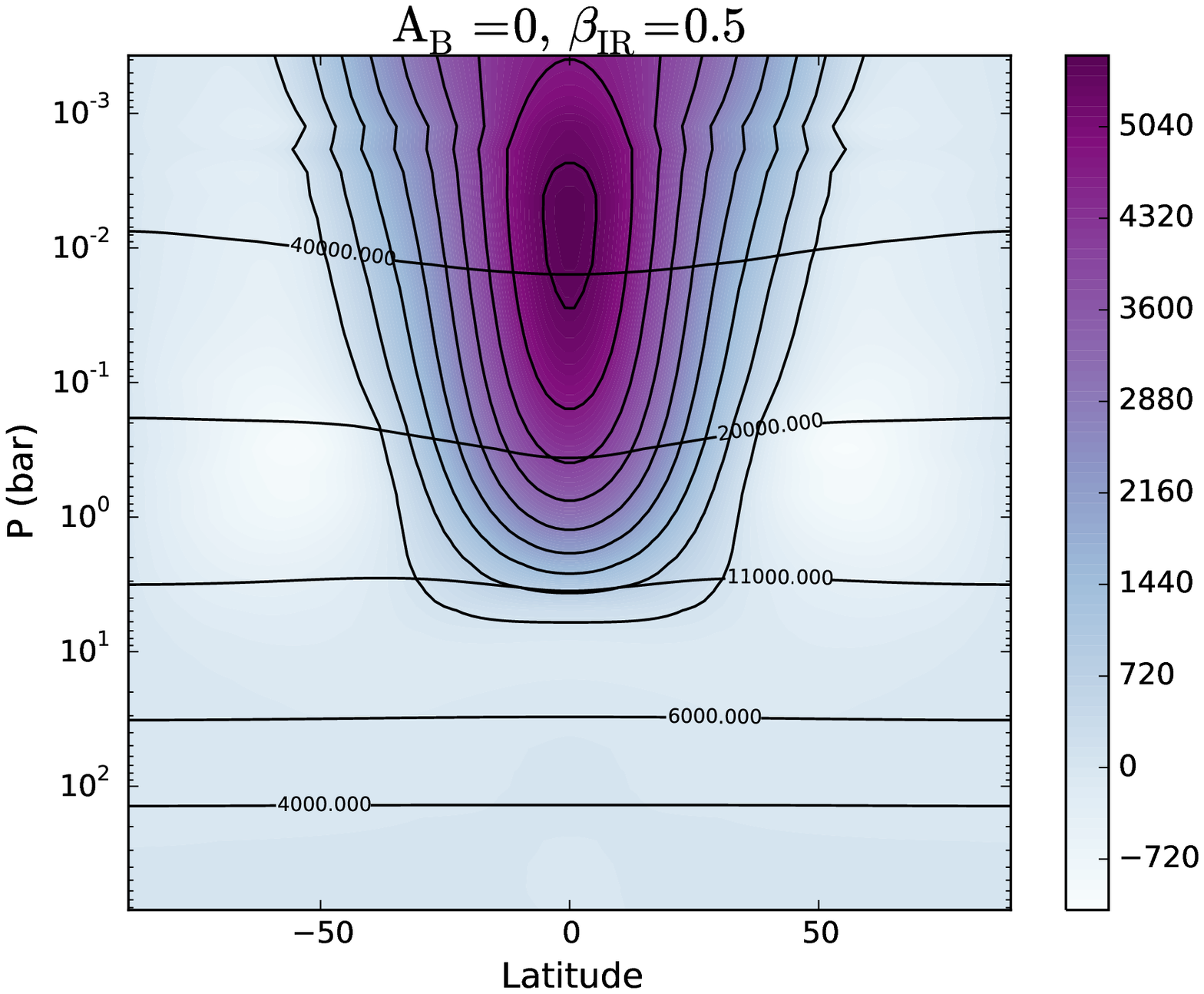}
\endminipage\hfill
\minipage{0.33\textwidth}
\includegraphics[width=\textwidth]{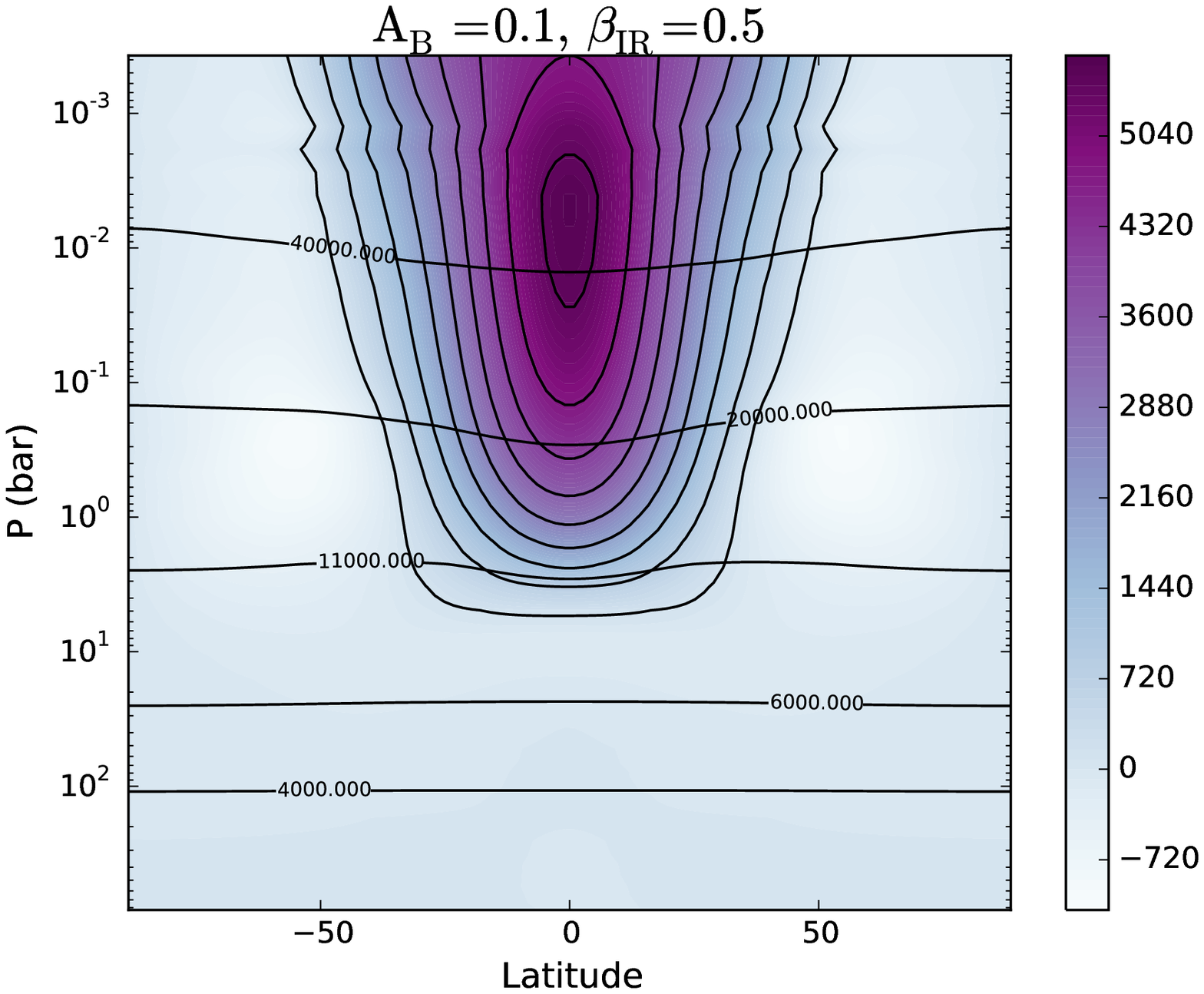}
\endminipage\hfill
\minipage{0.33\textwidth}
\includegraphics[width=\textwidth]{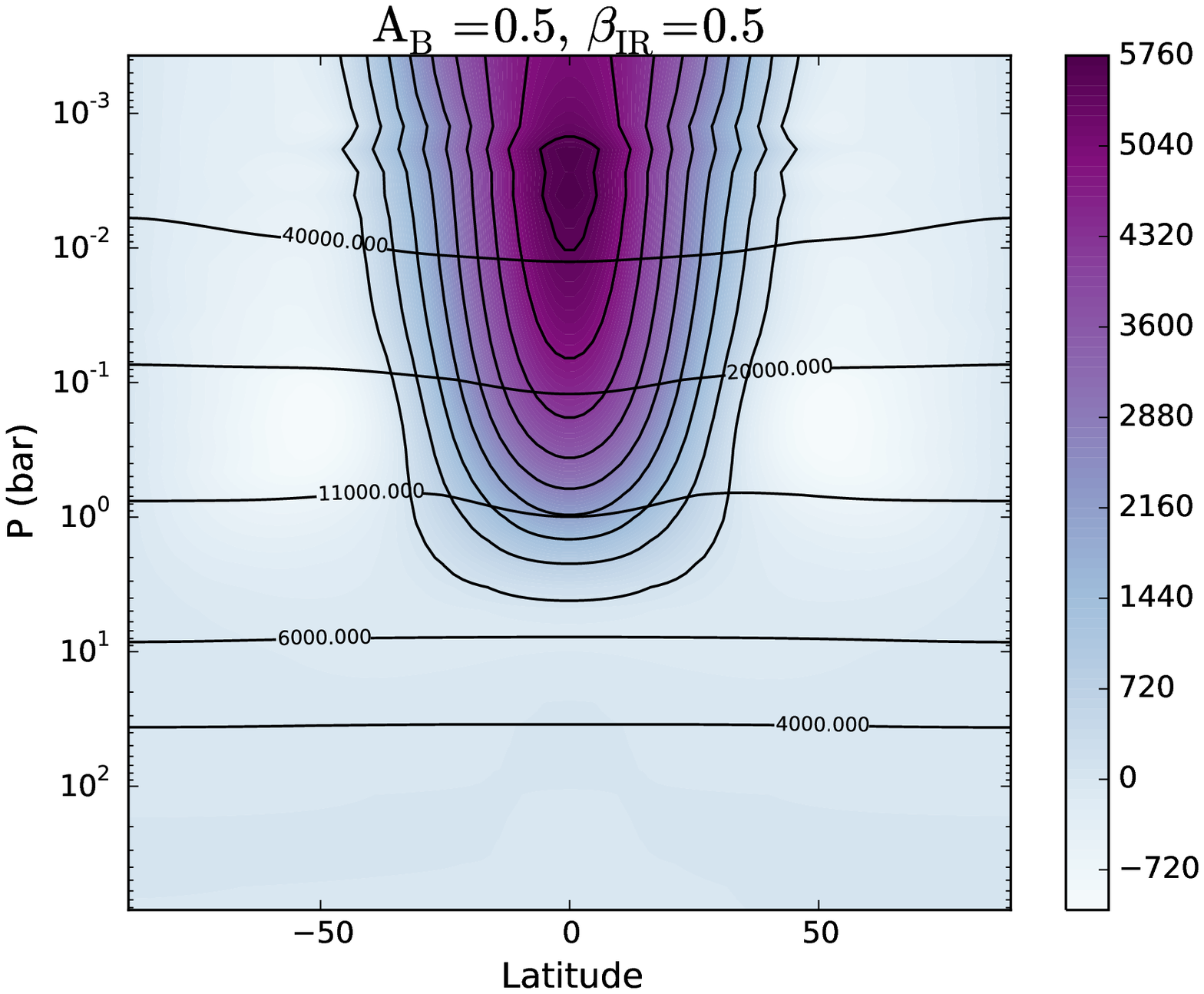}
\endminipage\hfill
\caption{Zonal-mean zonal wind of our suite of GCMs; physical units of the contours are in m s$^{-1}$.  The left, middle and right column are for $A_{\rm B}=0, 0.1$ and 0.5, respectively.  The top, middle and bottom rows are for $\beta_{\rm IR}=1, 0.75$ and 0.5, respectively.  Also overplotted are contours of the potential temperature in K.}
\label{fig:zonalwind}
\end{figure*}

\begin{figure*}
\minipage{0.33\textwidth}
\includegraphics[width=\textwidth]{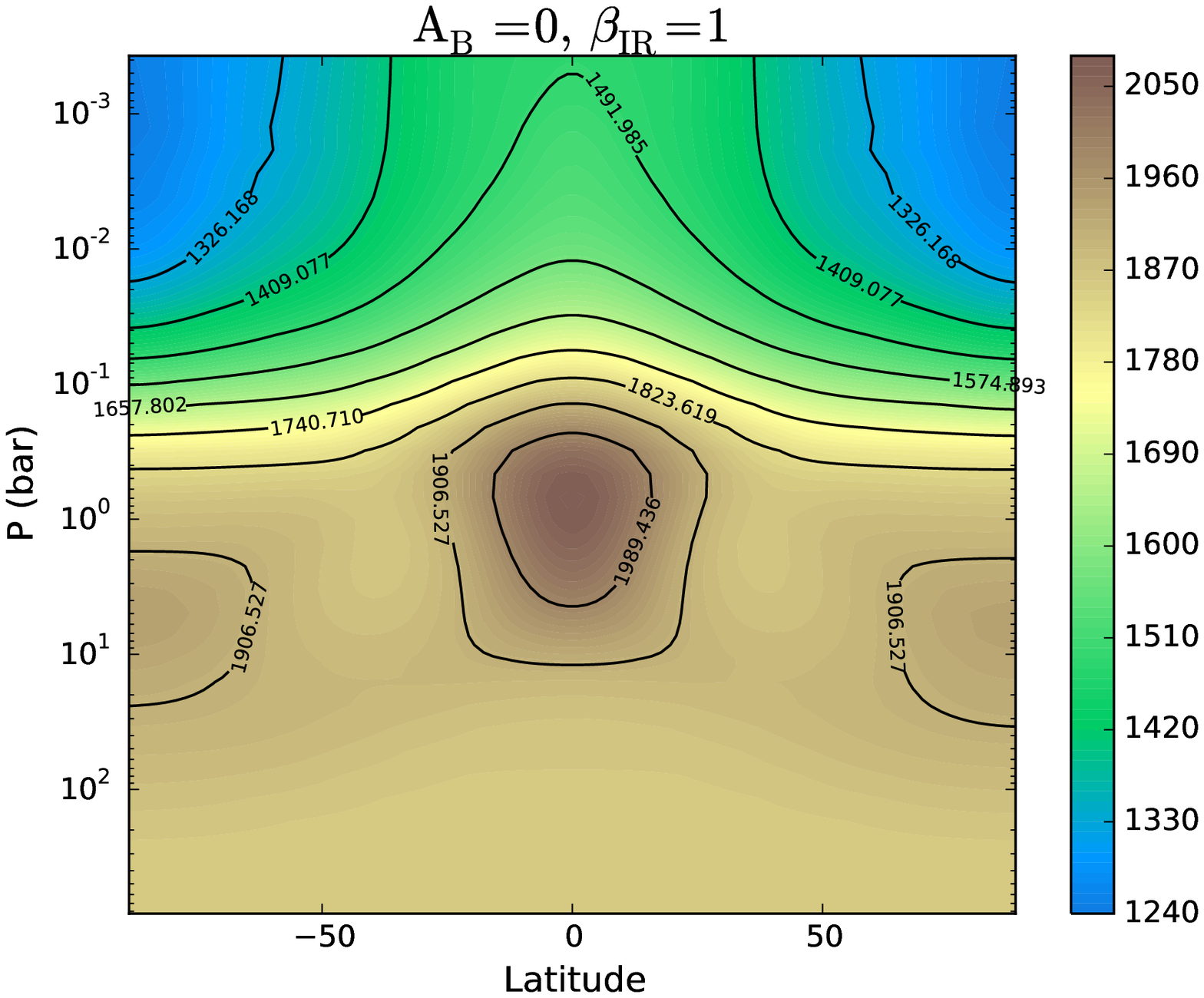}
\endminipage\hfill
\minipage{0.33\textwidth}
\includegraphics[width=\textwidth]{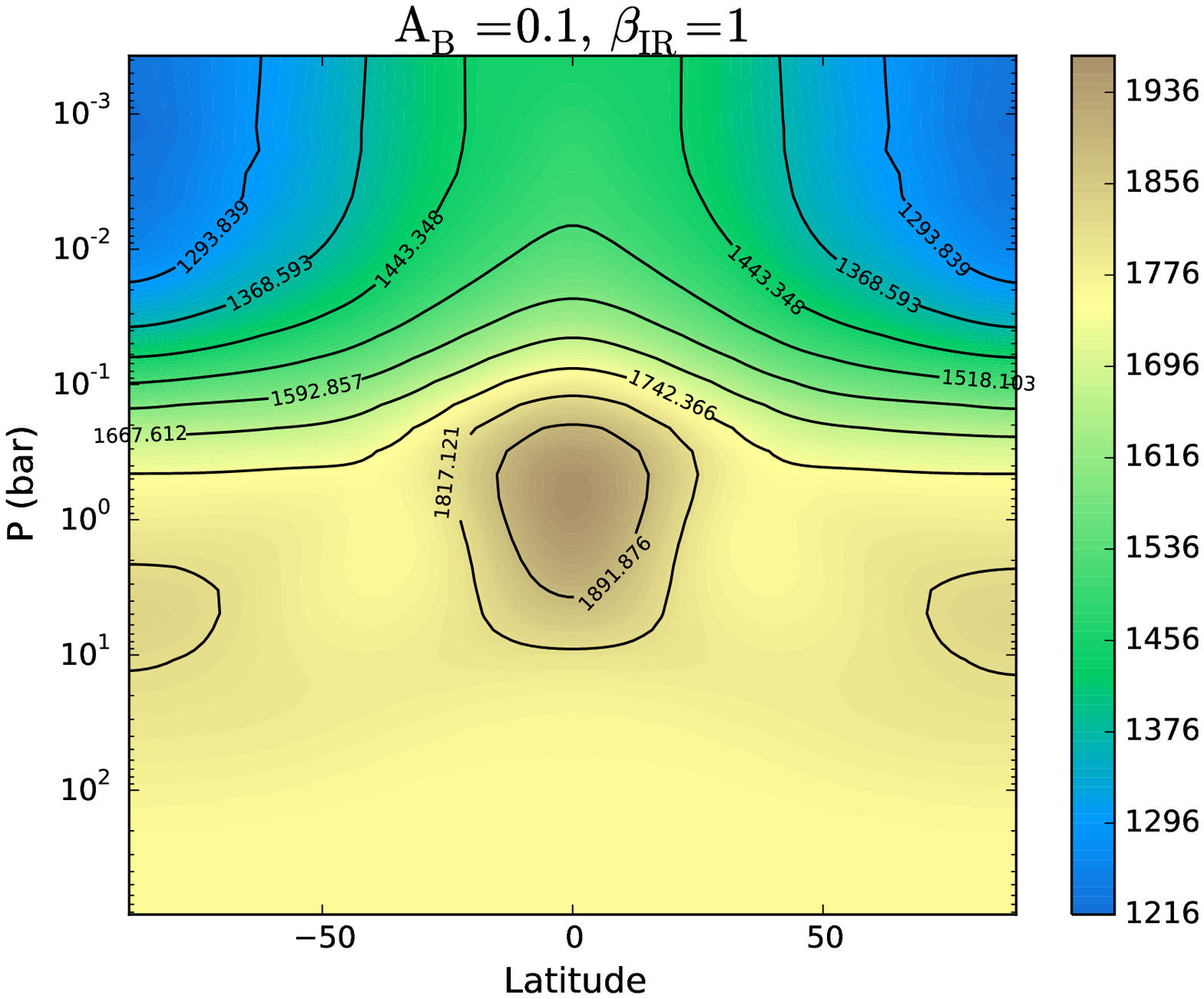}
\endminipage\hfill
\minipage{0.33\textwidth}
\includegraphics[width=\textwidth]{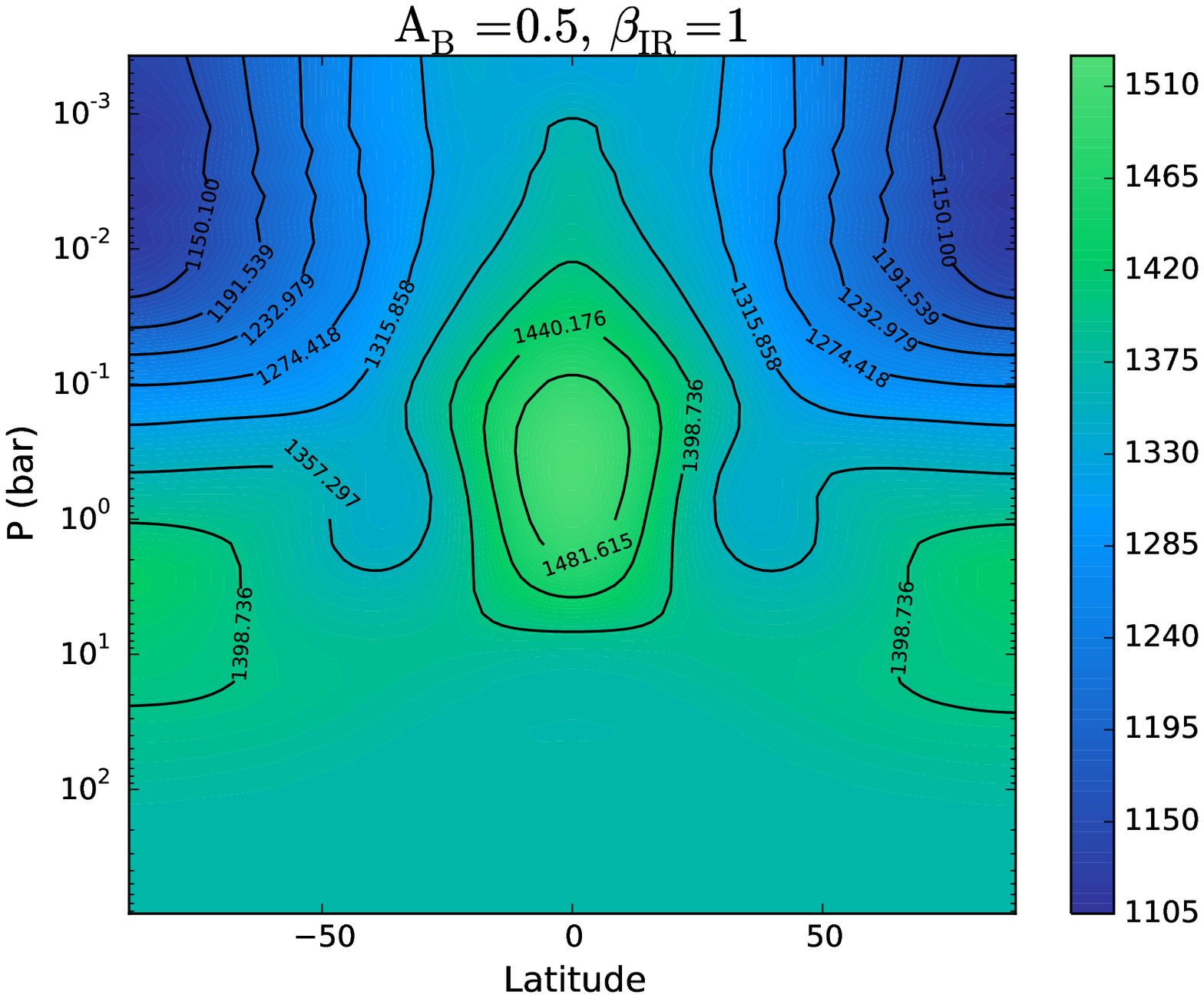}
\endminipage\hfill
\minipage{0.33\textwidth}
\includegraphics[width=\textwidth]{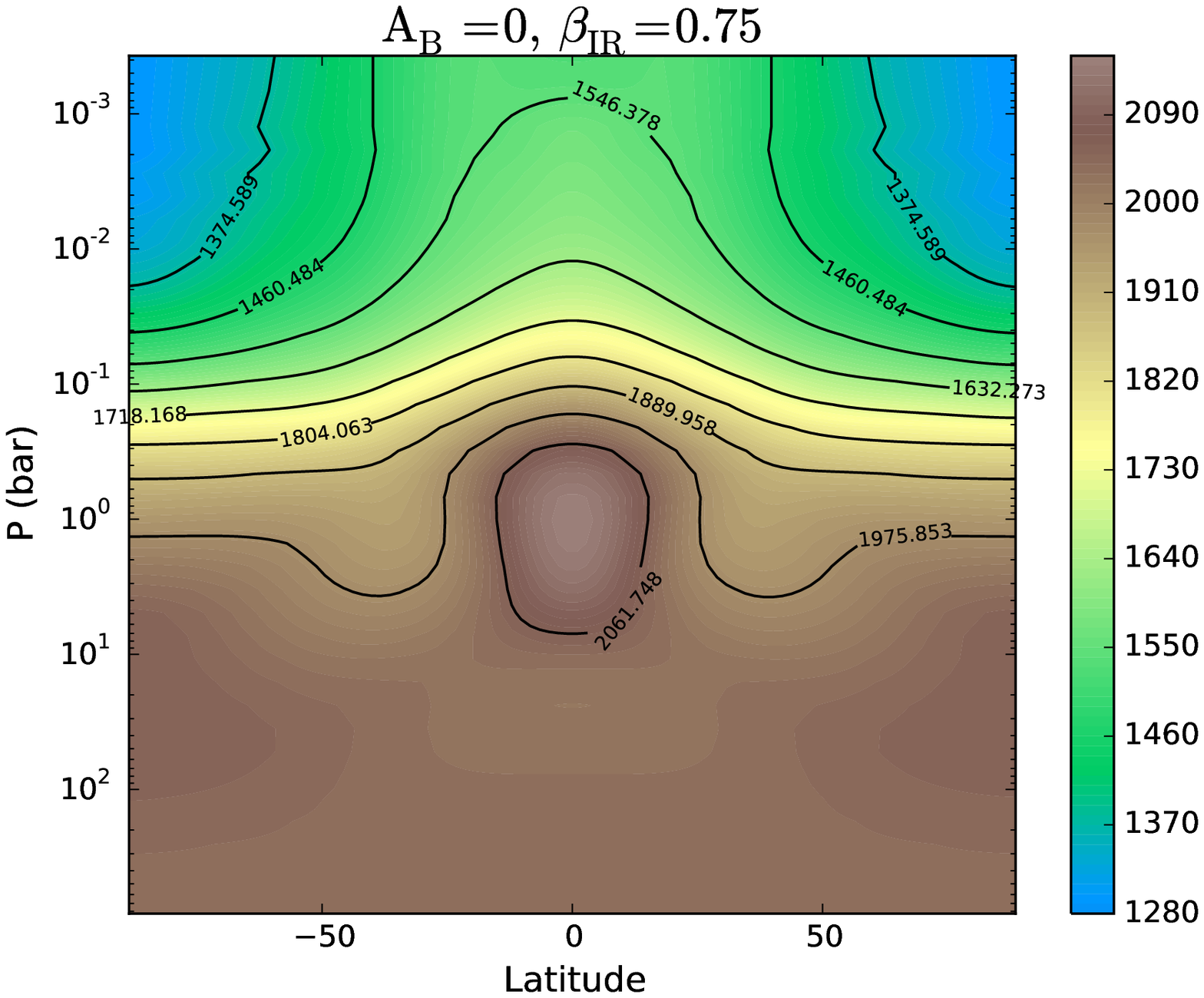}
\endminipage\hfill
\minipage{0.33\textwidth}
\includegraphics[width=\textwidth]{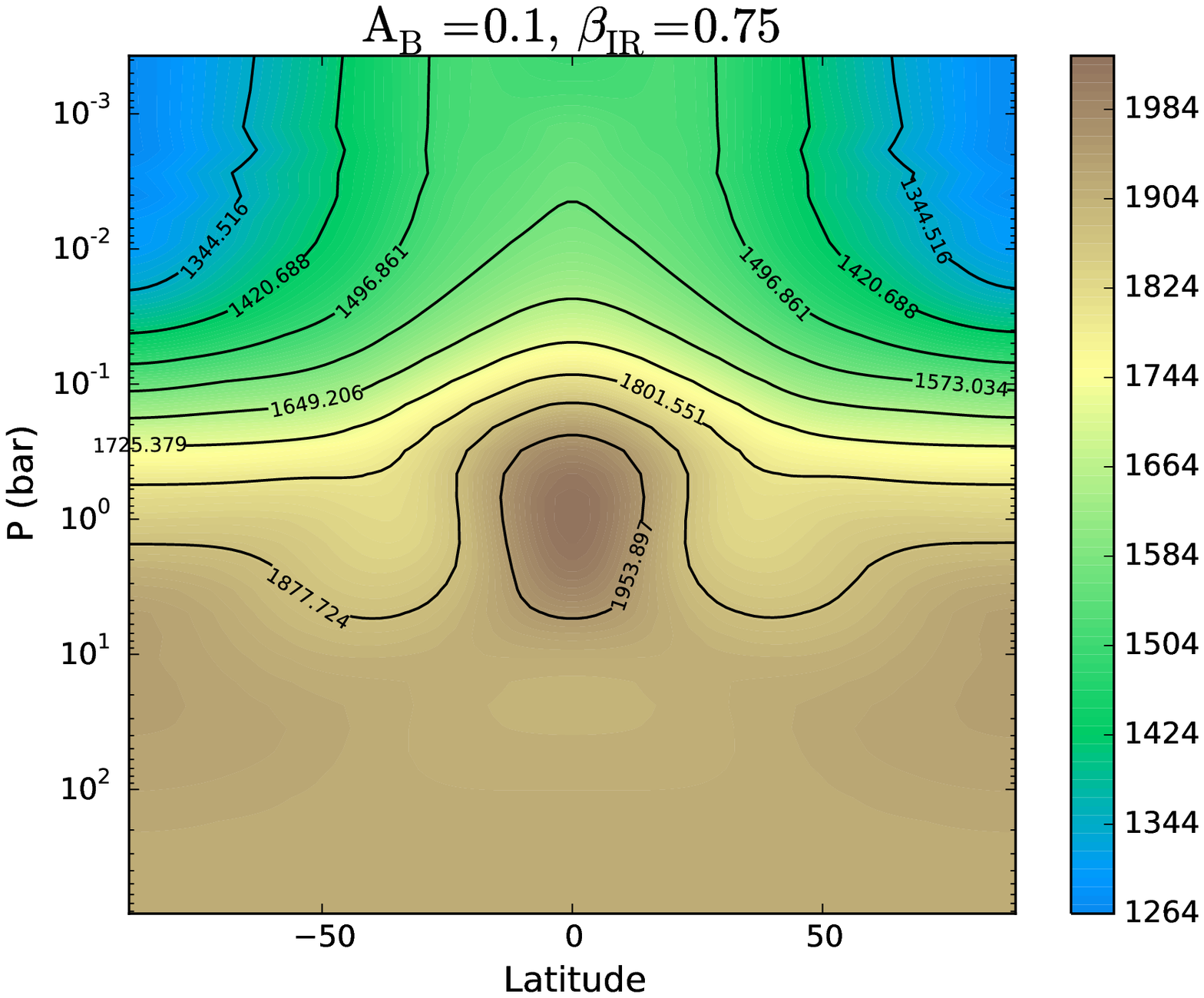}
\endminipage\hfill
\minipage{0.33\textwidth}
\includegraphics[width=\textwidth]{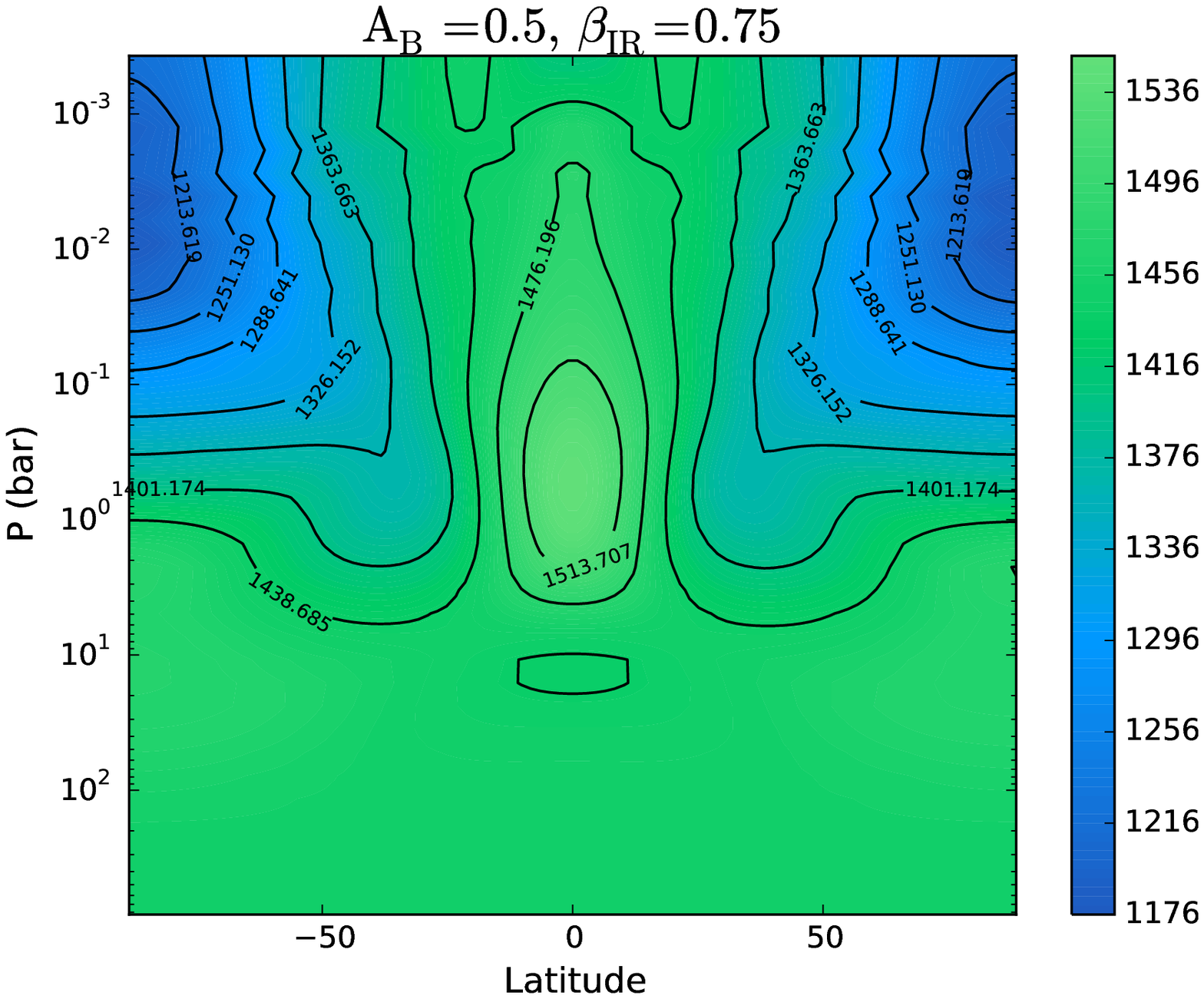}
\endminipage\hfill
\minipage{0.33\textwidth}
\includegraphics[width=\textwidth]{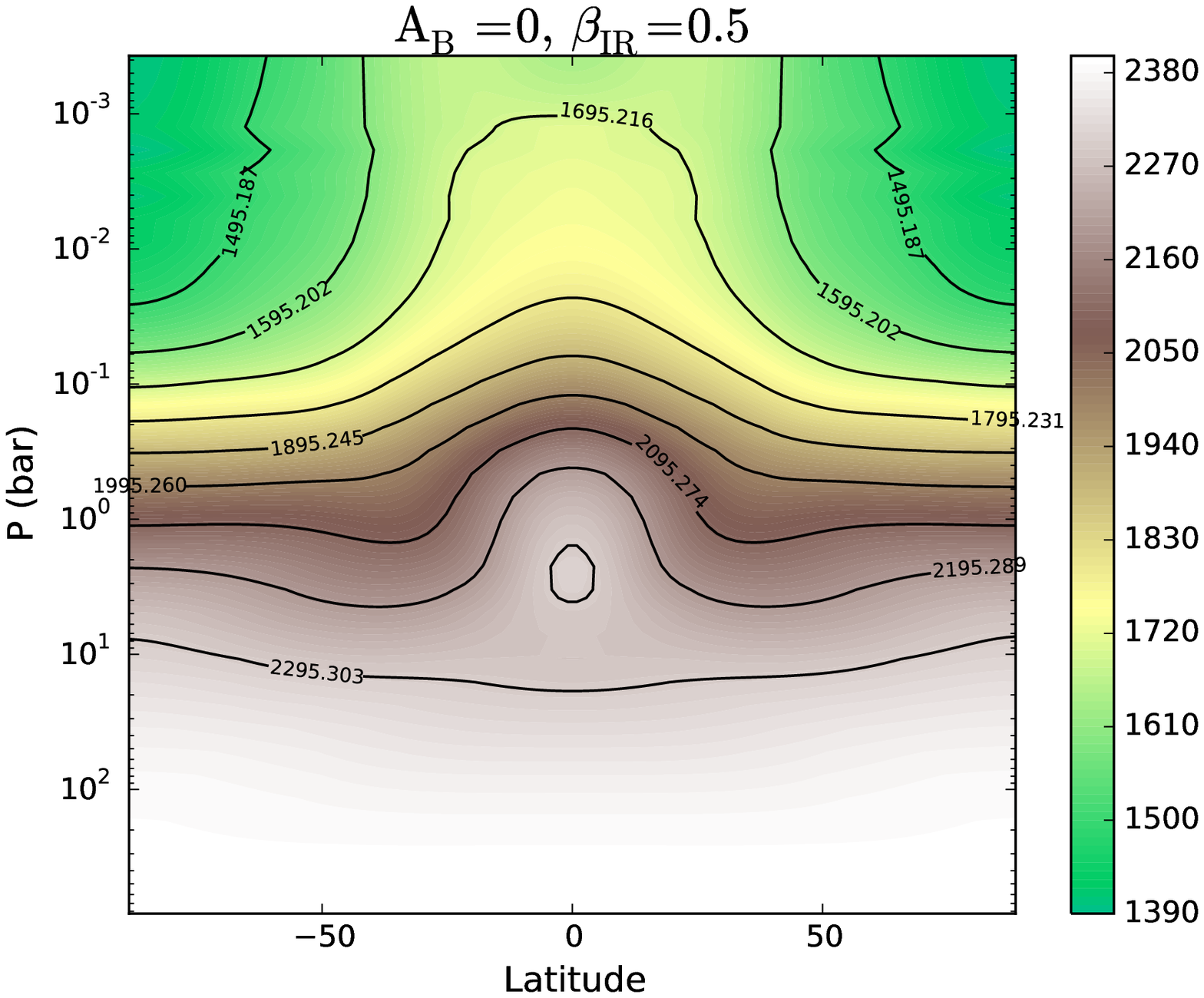}
\endminipage\hfill
\minipage{0.33\textwidth}
\includegraphics[width=\textwidth]{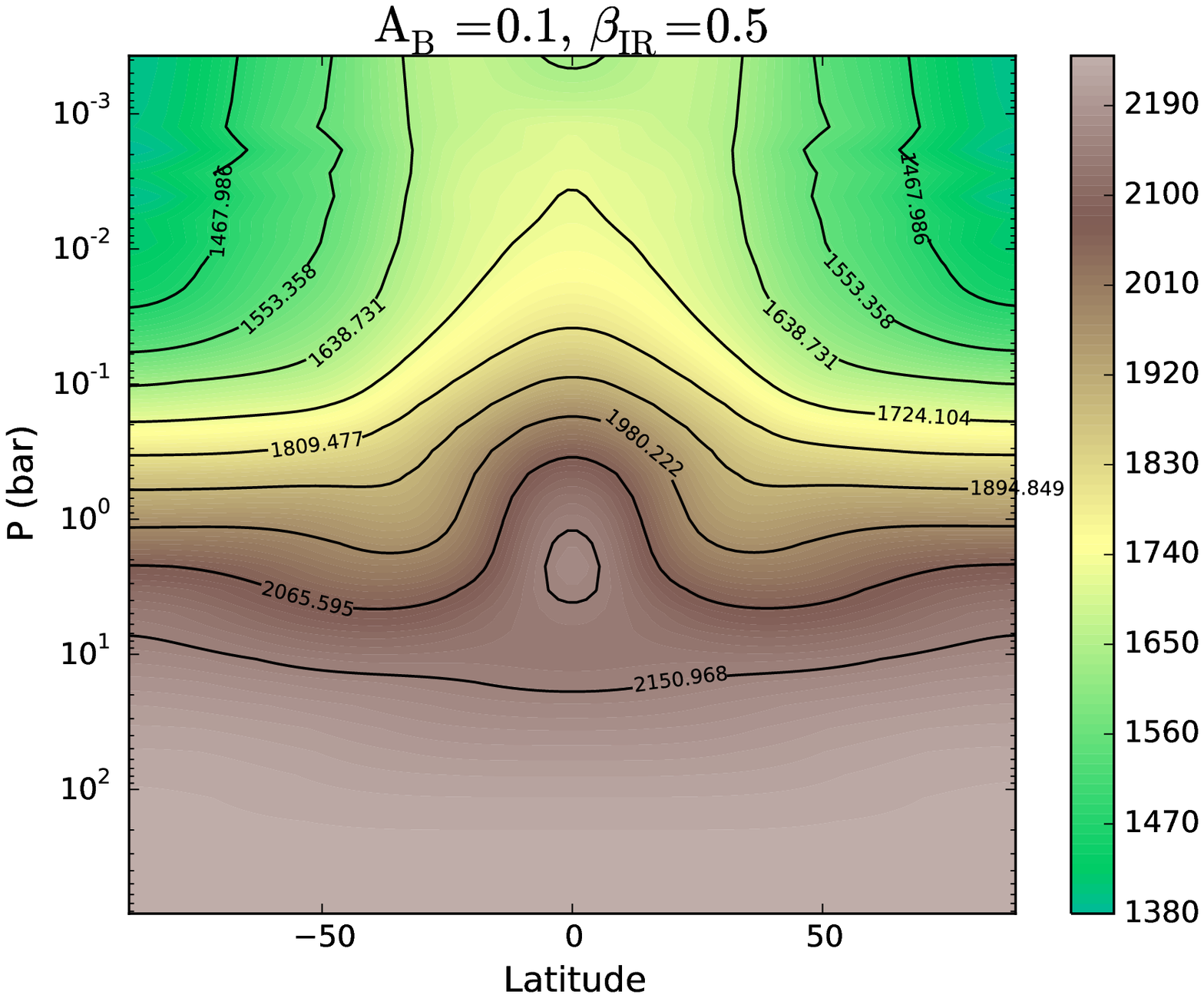}
\endminipage\hfill
\minipage{0.33\textwidth}
\includegraphics[width=\textwidth]{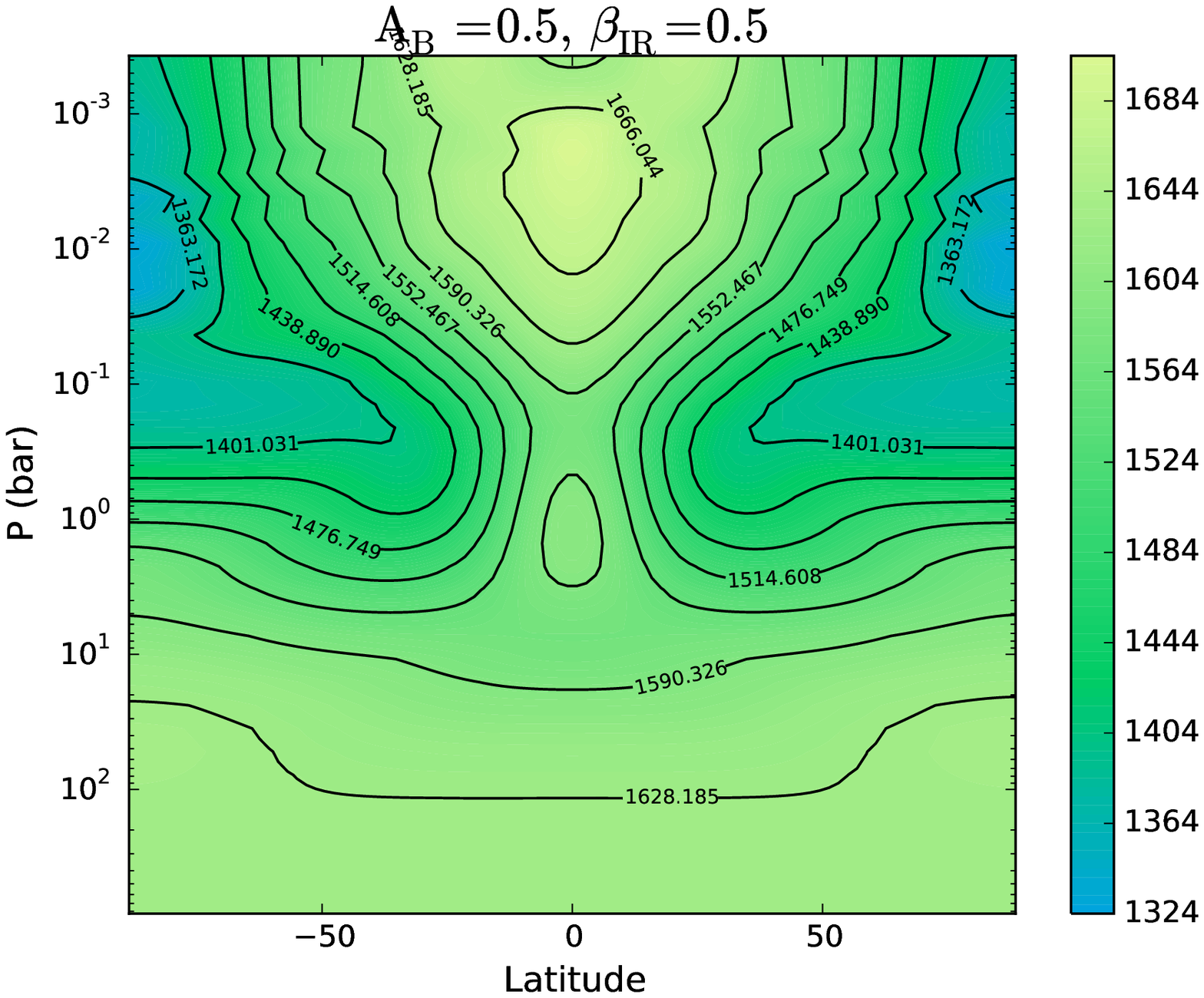}
\endminipage\hfill
\caption{Same as Figure \ref{fig:zonalwind}, but for the contours of the zonal-mean temperature in physical units of K.}
\label{fig:temperature}
\end{figure*}

\begin{figure*}
\minipage{0.33\textwidth}
\includegraphics[width=\textwidth]{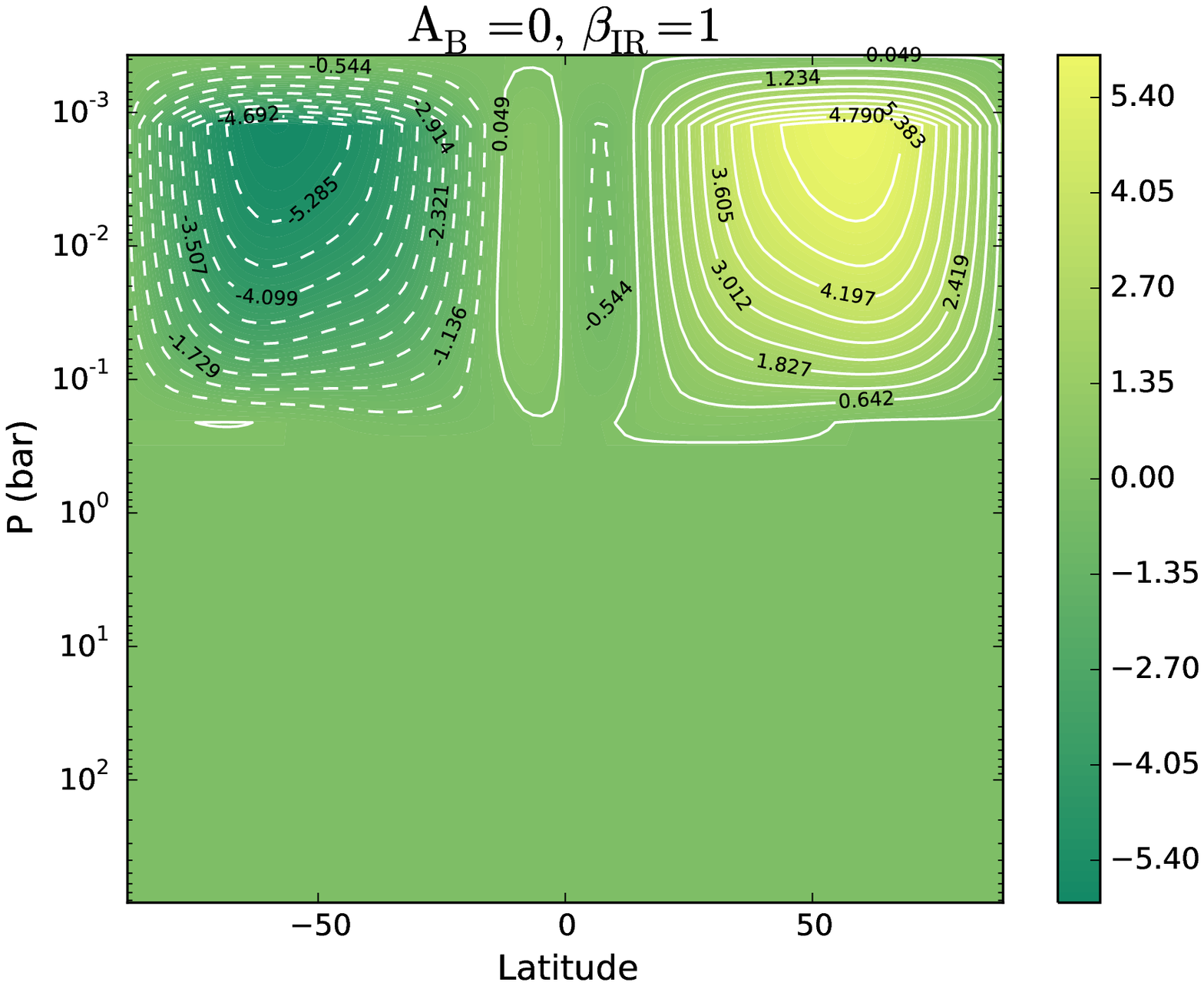}
\endminipage\hfill
\minipage{0.33\textwidth}
\includegraphics[width=\textwidth]{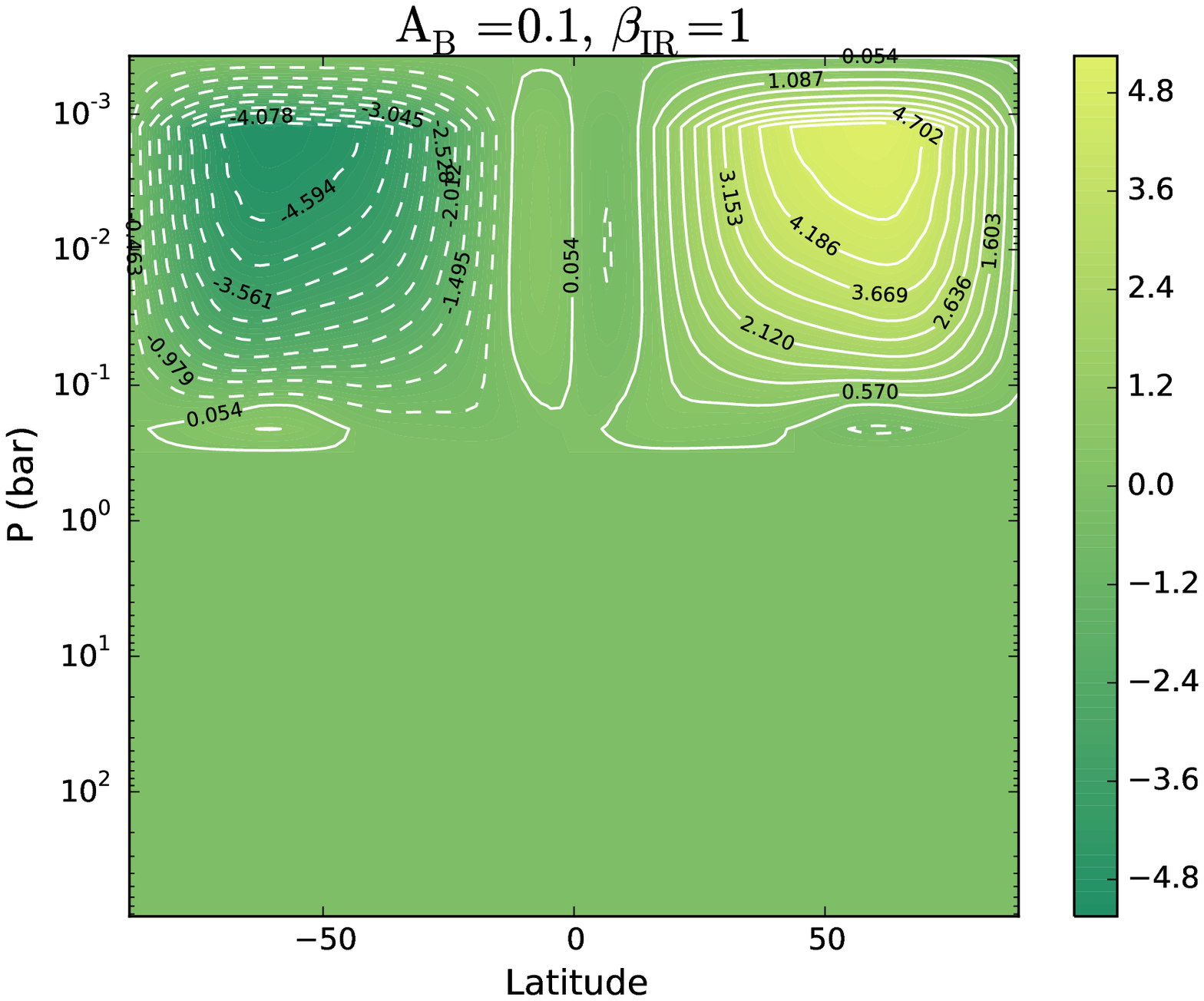}
\endminipage\hfill
\minipage{0.33\textwidth}
\includegraphics[width=\textwidth]{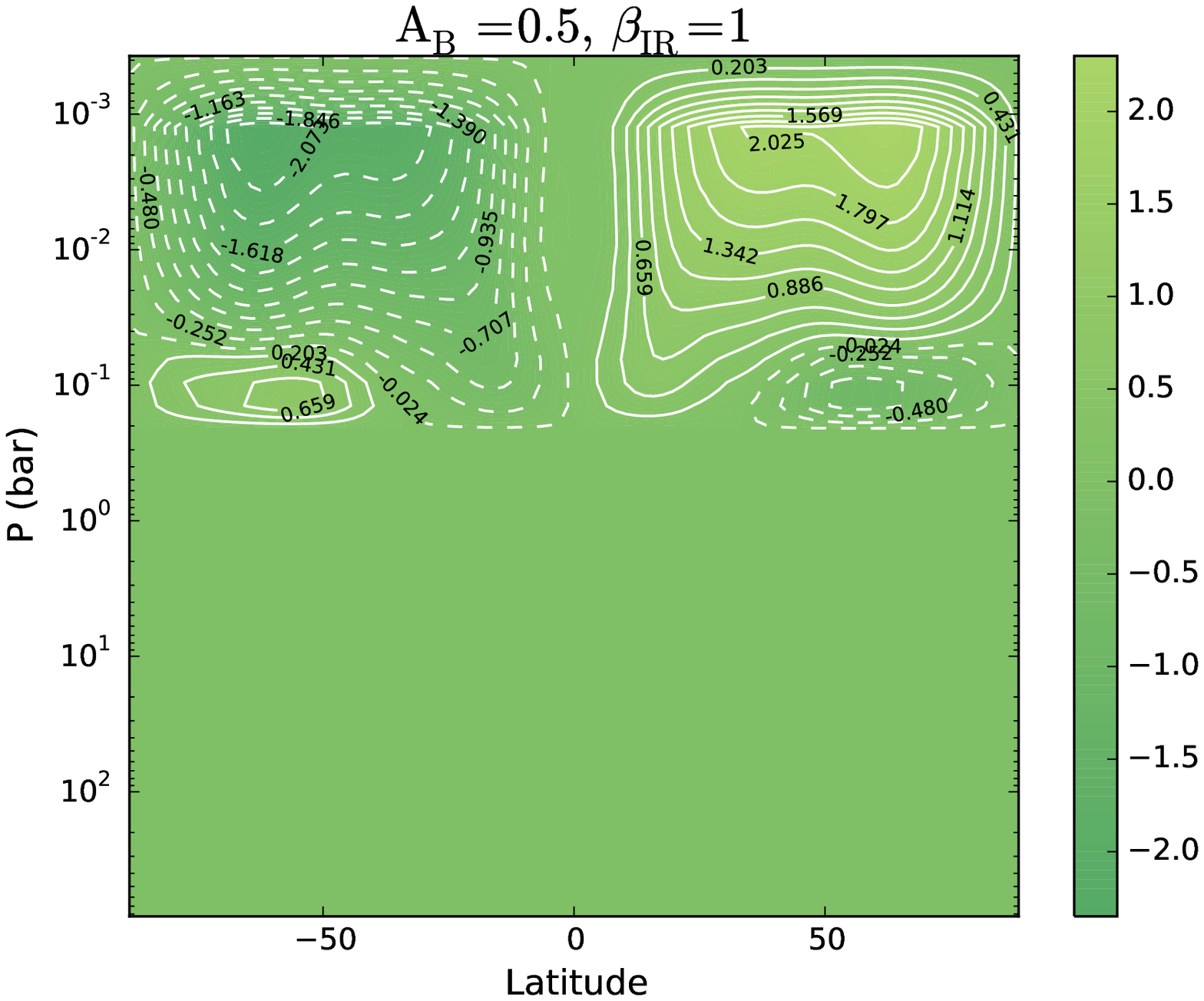}
\endminipage\hfill
\minipage{0.33\textwidth}
\includegraphics[width=\textwidth]{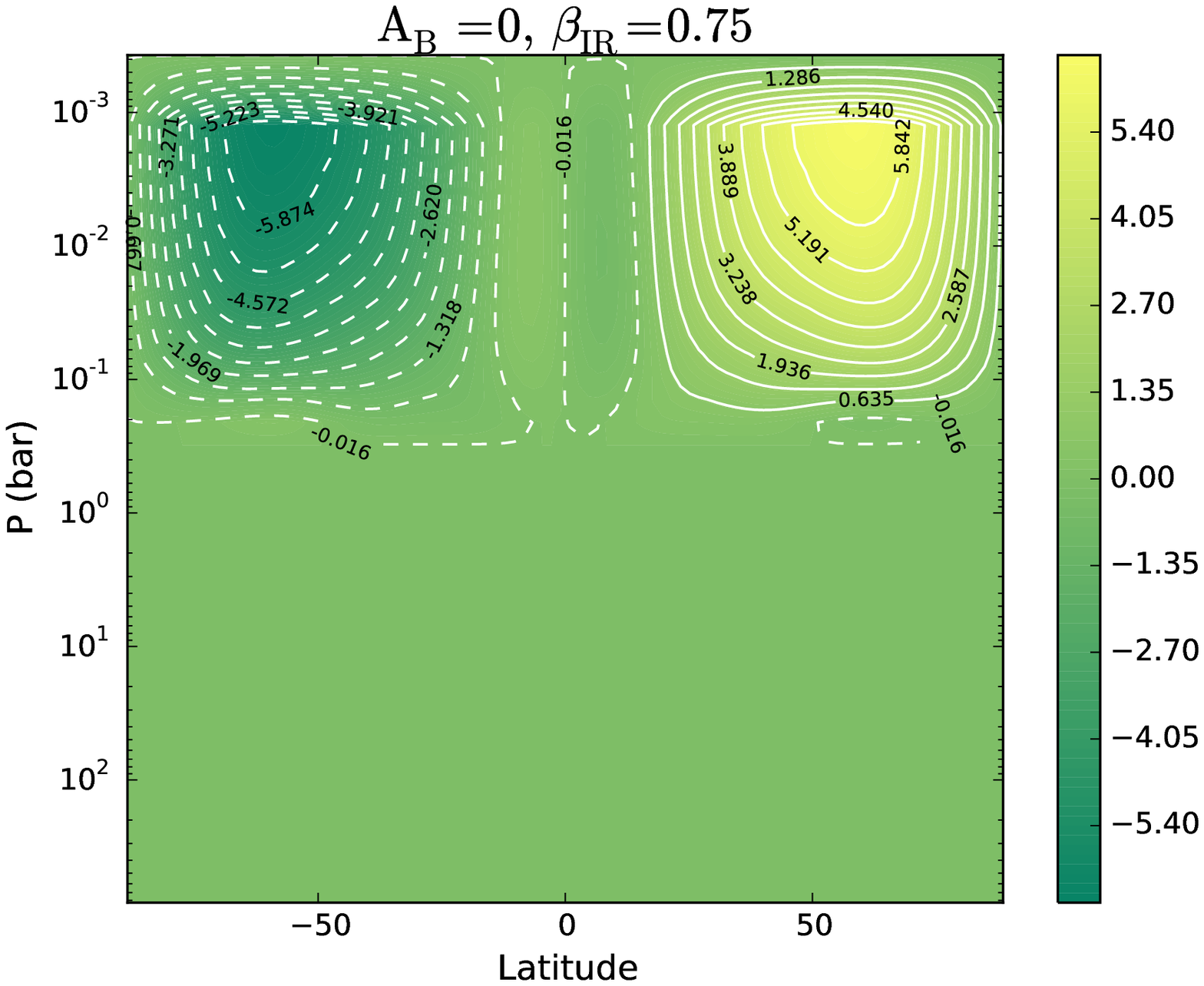}
\endminipage\hfill
\minipage{0.33\textwidth}
\includegraphics[width=\textwidth]{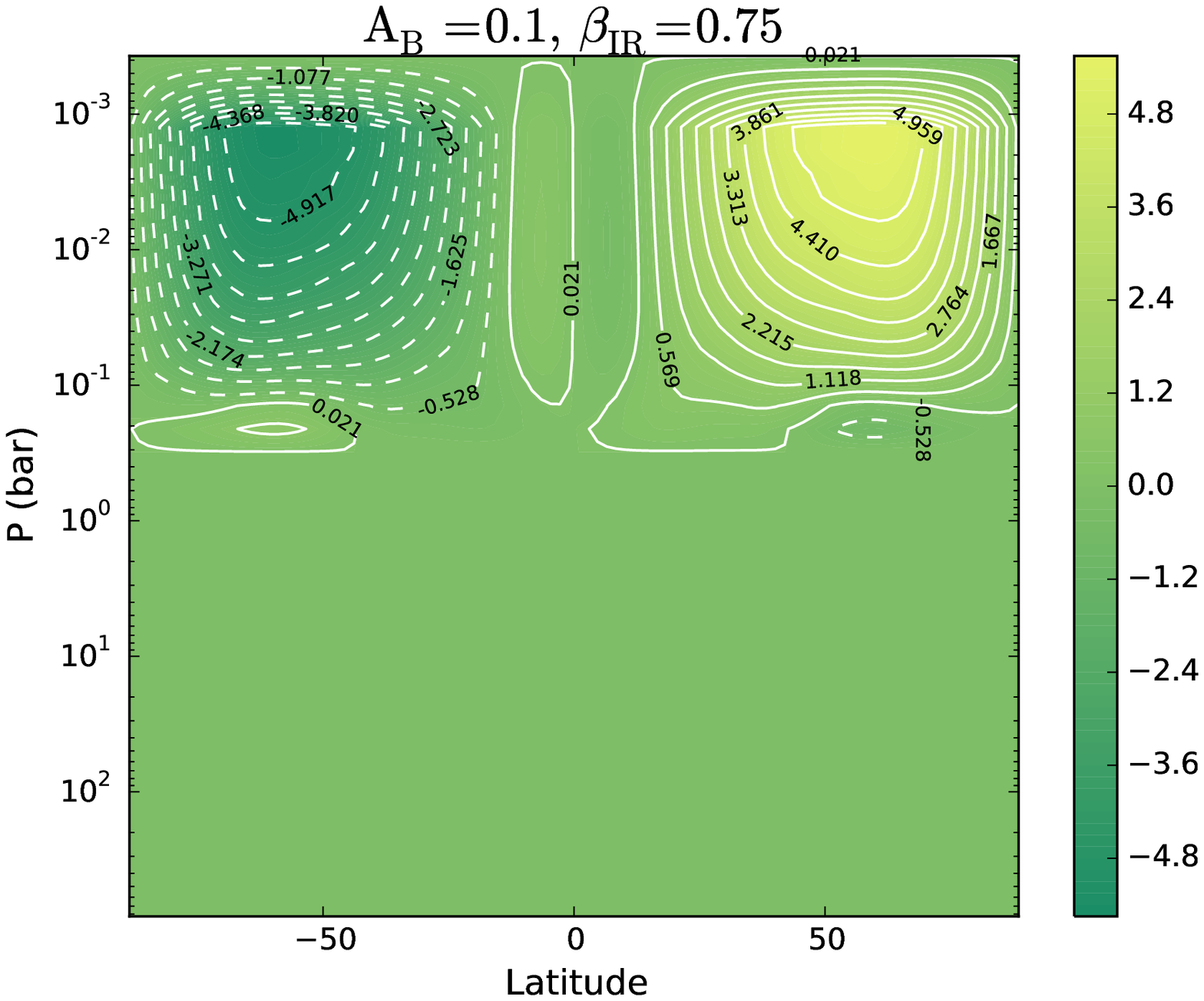}
\endminipage\hfill
\minipage{0.33\textwidth}
\includegraphics[width=\textwidth]{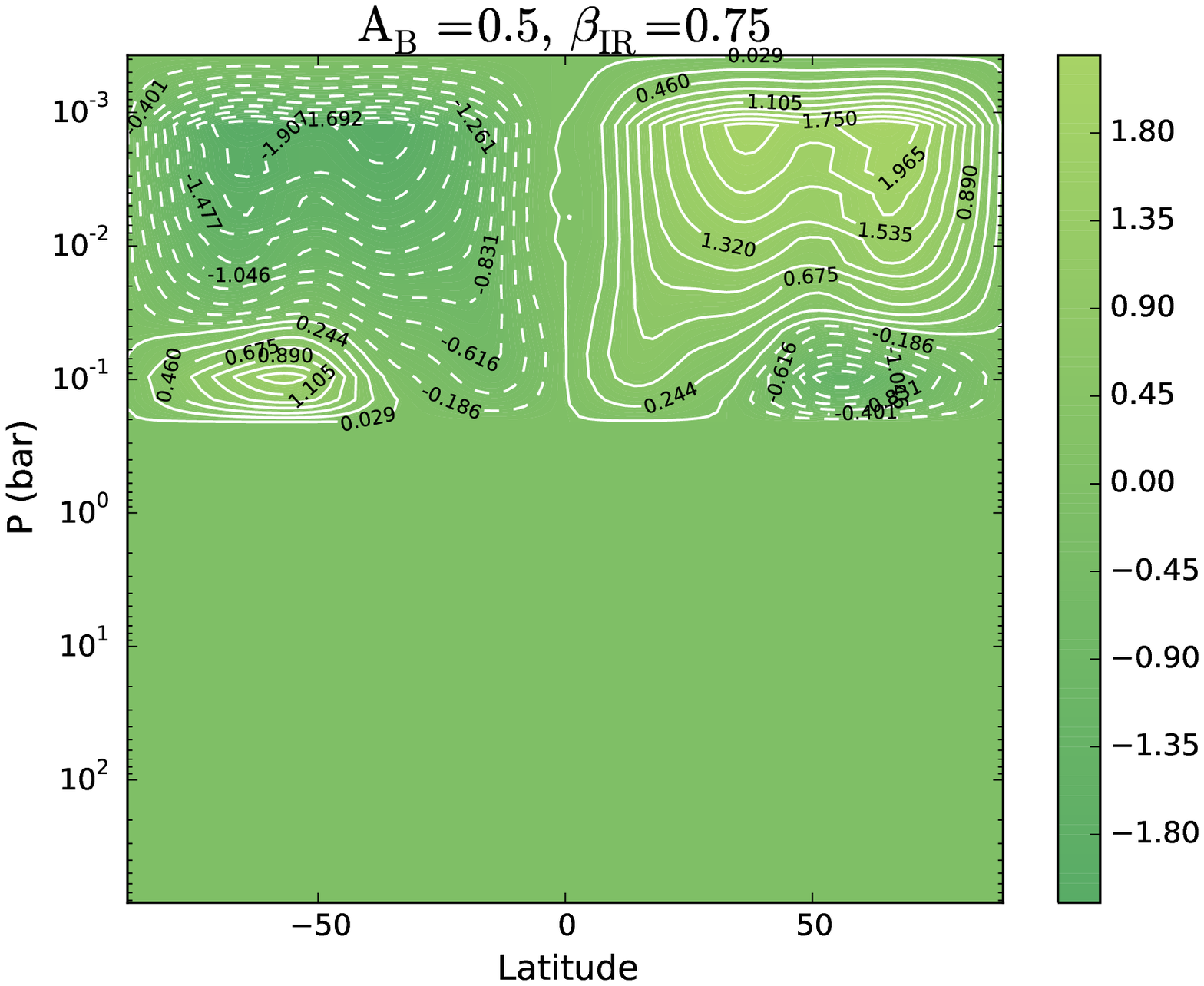}
\endminipage\hfill
\minipage{0.33\textwidth}
\includegraphics[width=\textwidth]{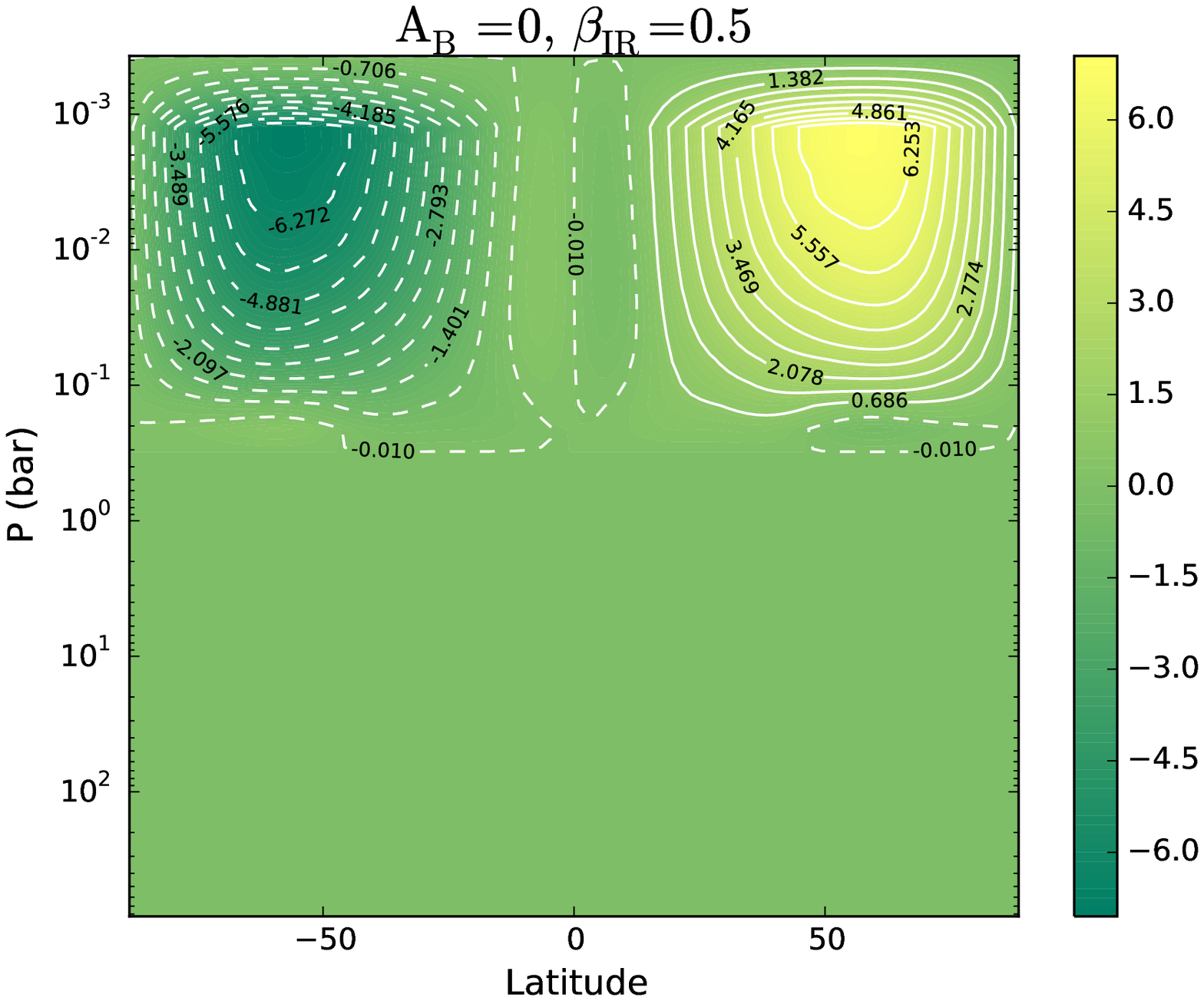}
\endminipage\hfill
\minipage{0.33\textwidth}
\includegraphics[width=\textwidth]{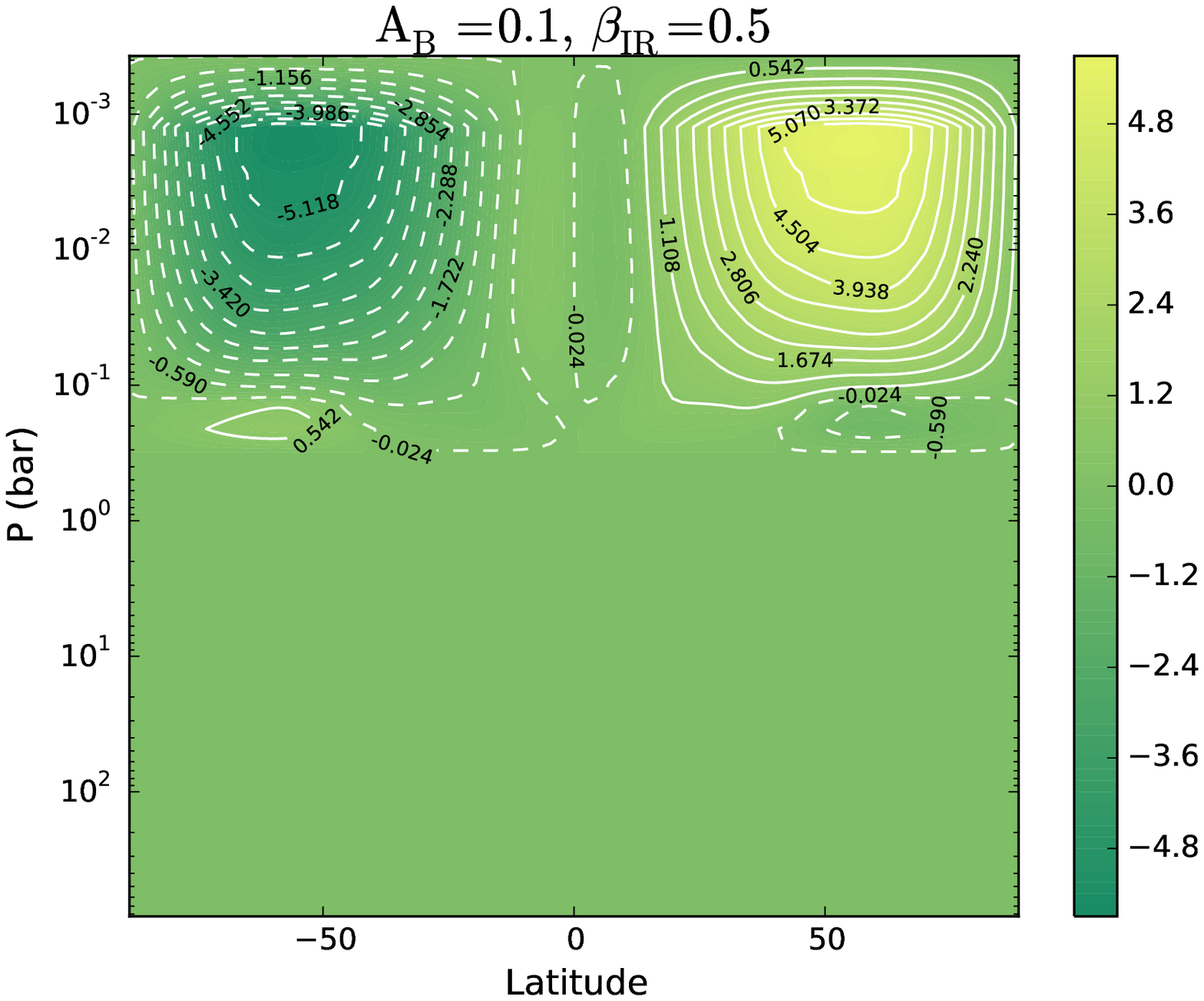}
\endminipage\hfill
\minipage{0.33\textwidth}
\includegraphics[width=\textwidth]{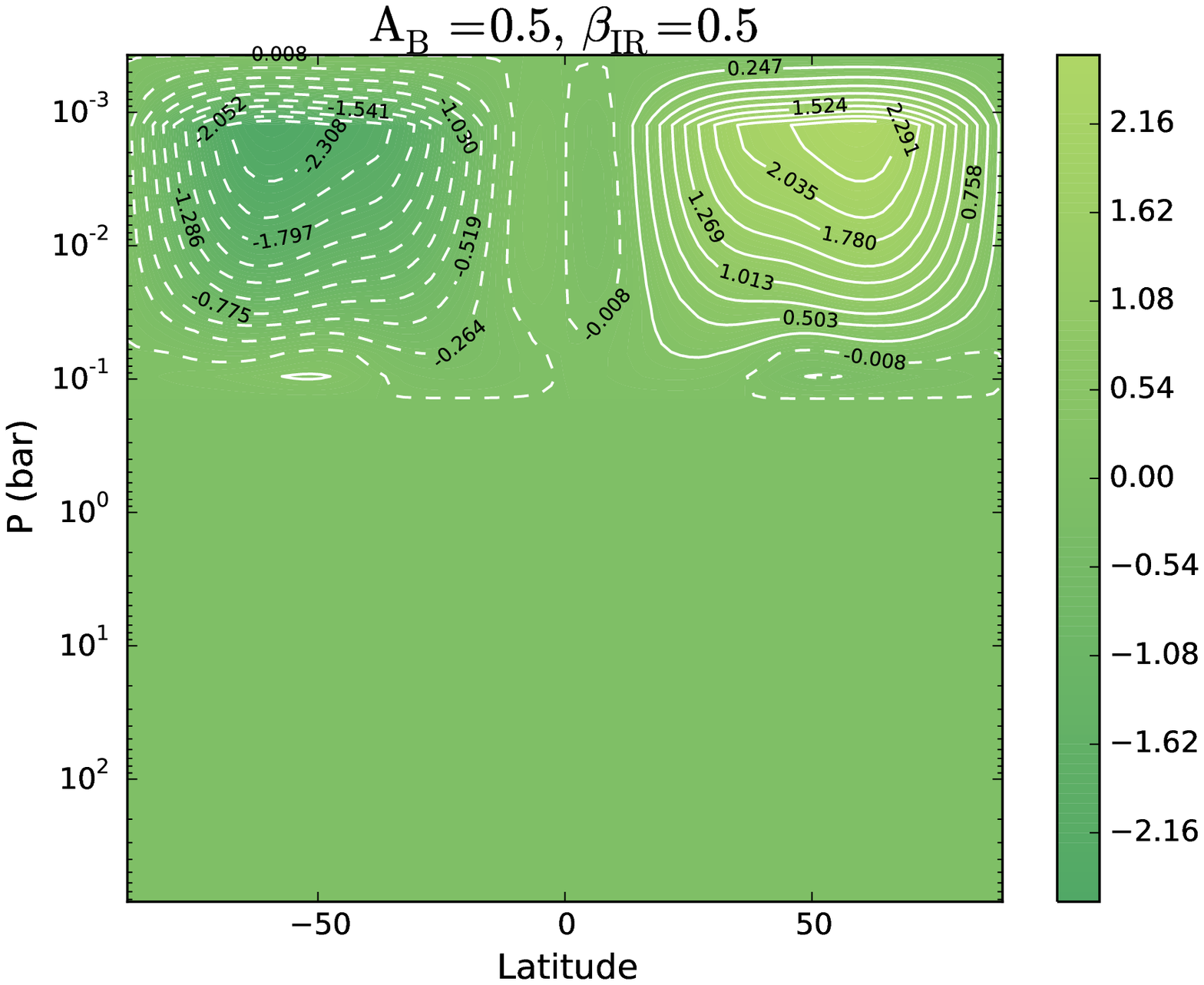}
\endminipage\hfill
\caption{Same as Figure \ref{fig:zonalwind}, but for the Euler-mean streamfunction of the dayside in physical units of $10^{13}$ kg s$^{-1}$.}
\label{fig:streamfunction}
\end{figure*}

\begin{figure*}
\minipage{0.33\textwidth}
\includegraphics[width=\textwidth]{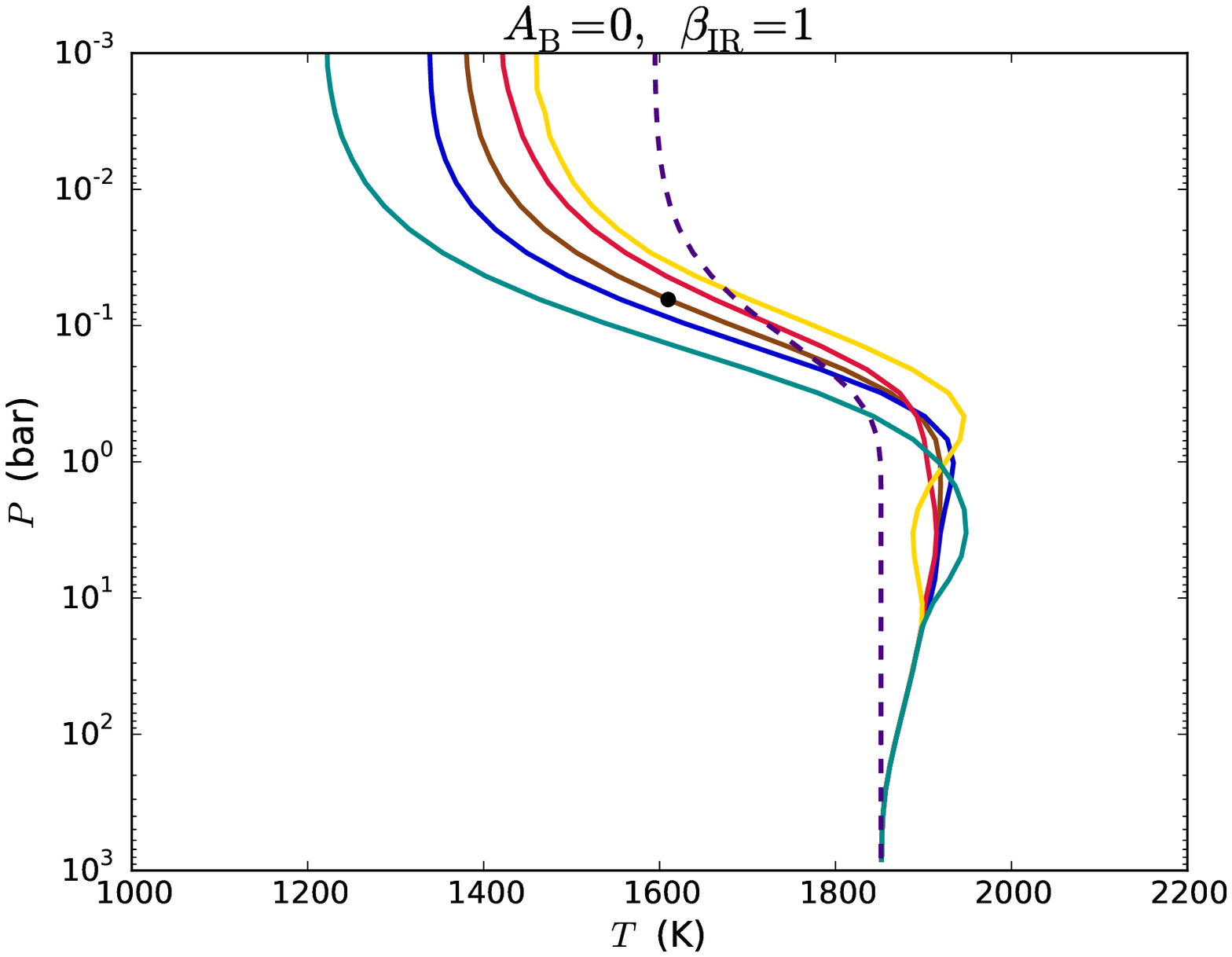}
\endminipage\hfill
\minipage{0.33\textwidth}
\includegraphics[width=\textwidth]{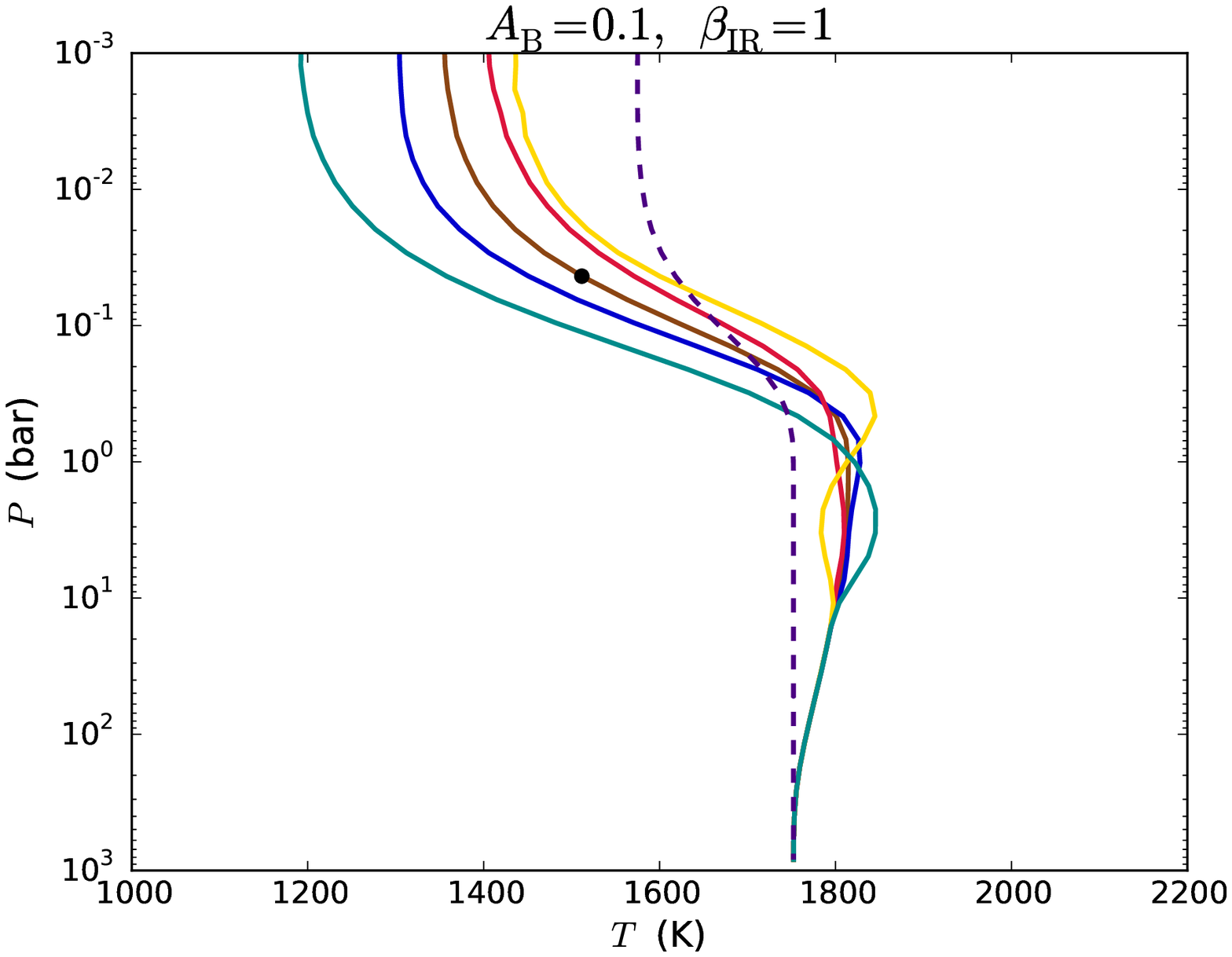}
\endminipage\hfill
\minipage{0.33\textwidth}
\includegraphics[width=\textwidth]{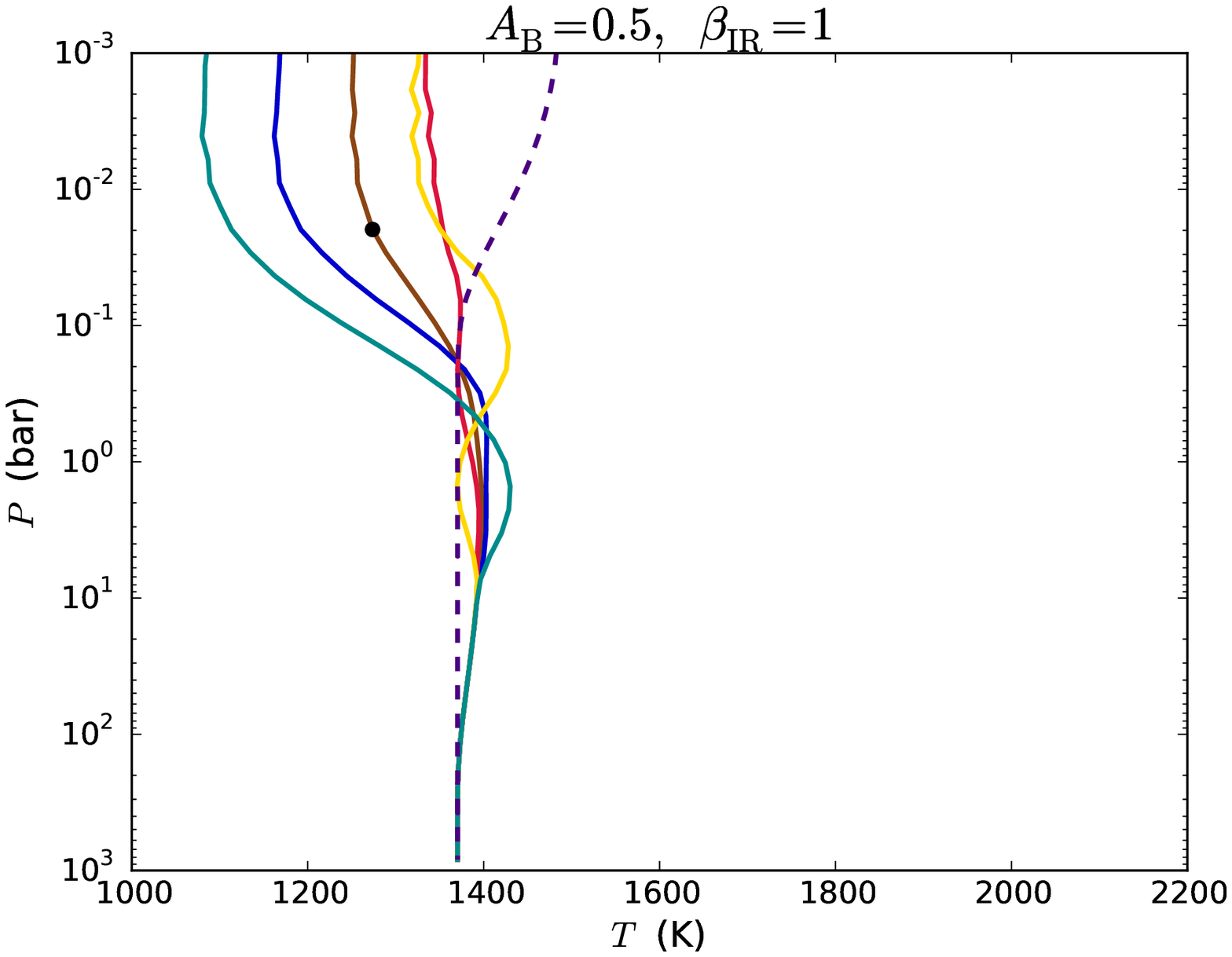}
\endminipage\hfill
\minipage{0.33\textwidth}
\includegraphics[width=\textwidth]{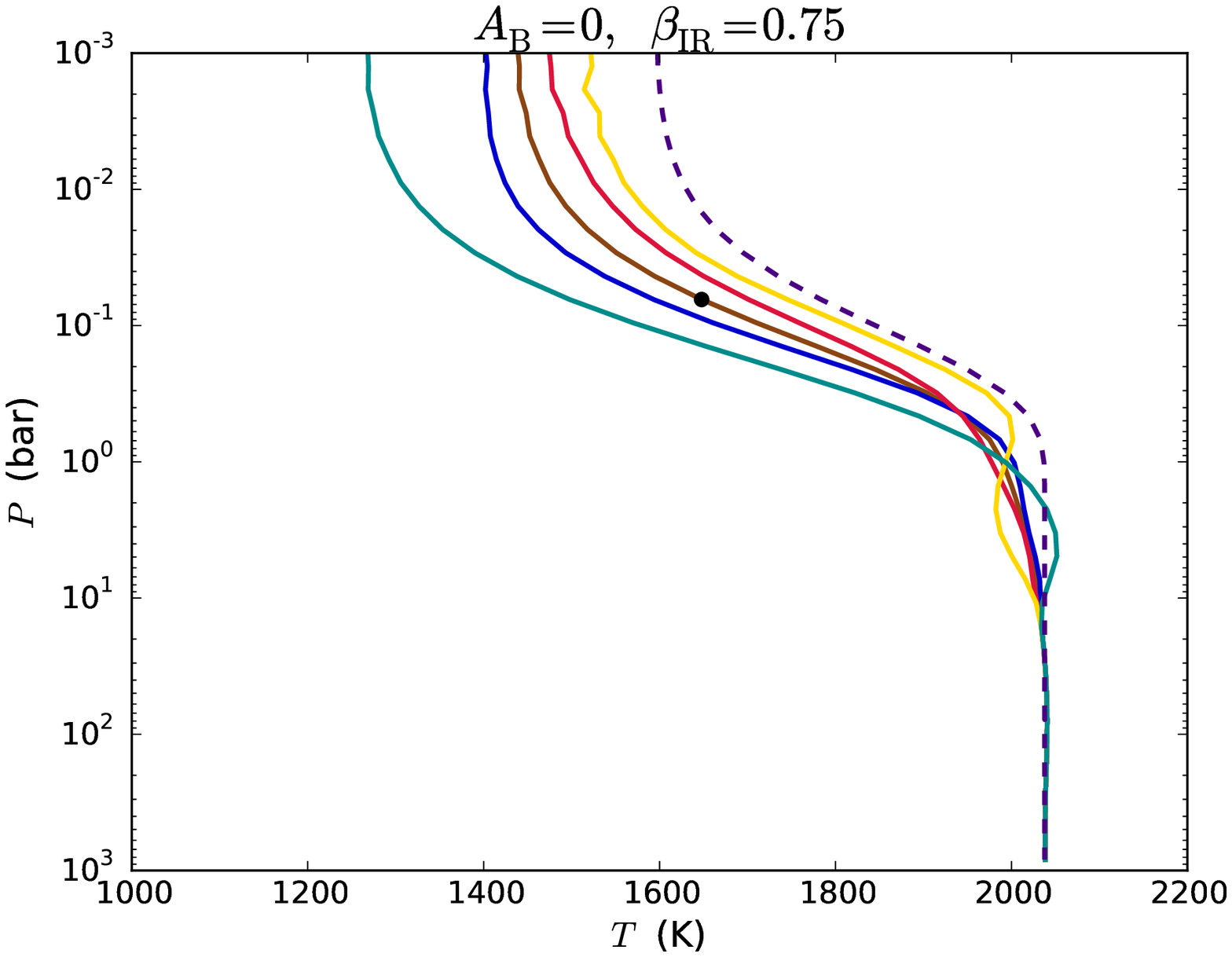}
\endminipage\hfill
\minipage{0.33\textwidth}
\includegraphics[width=\textwidth]{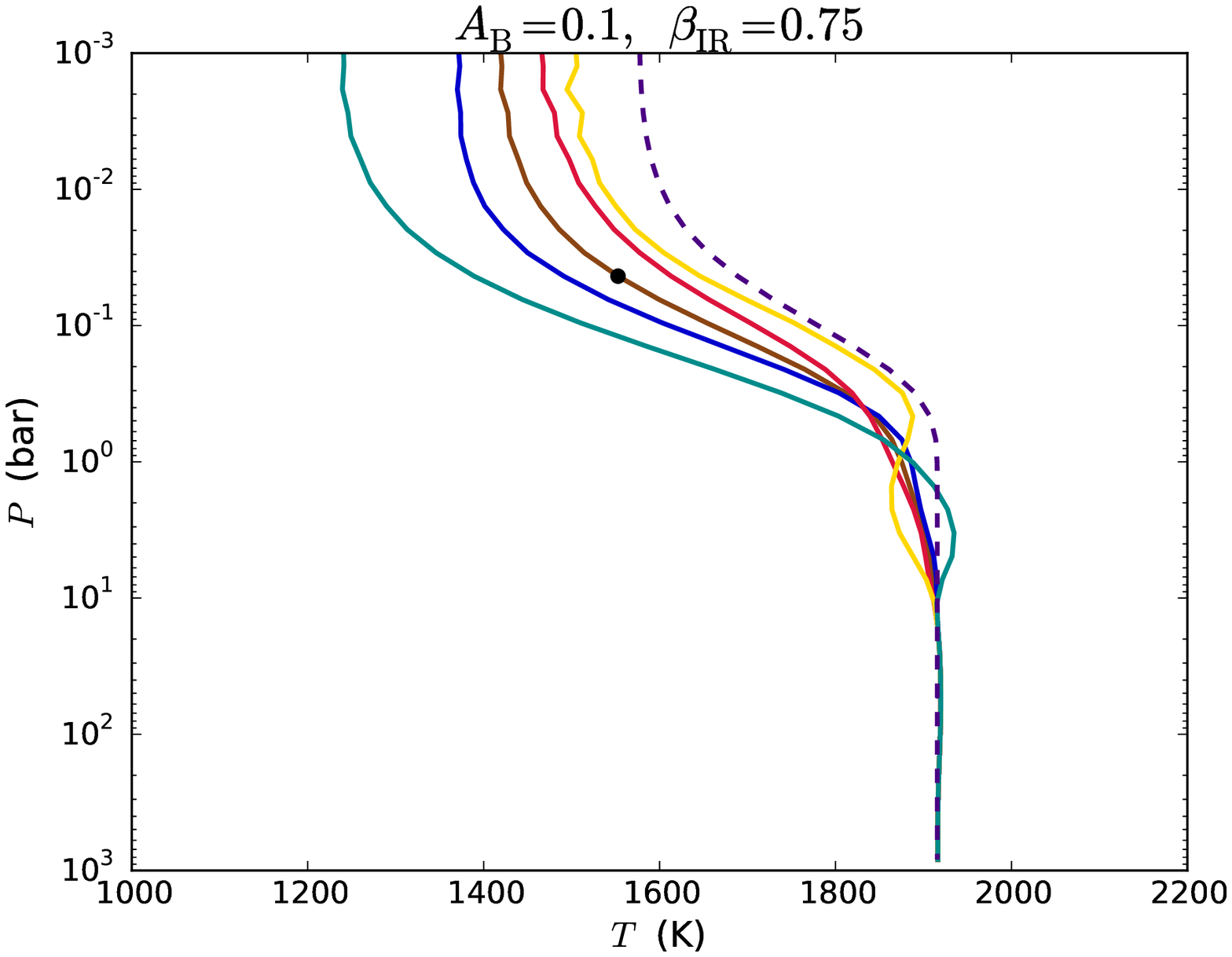}
\endminipage\hfill
\minipage{0.33\textwidth}
\includegraphics[width=\textwidth]{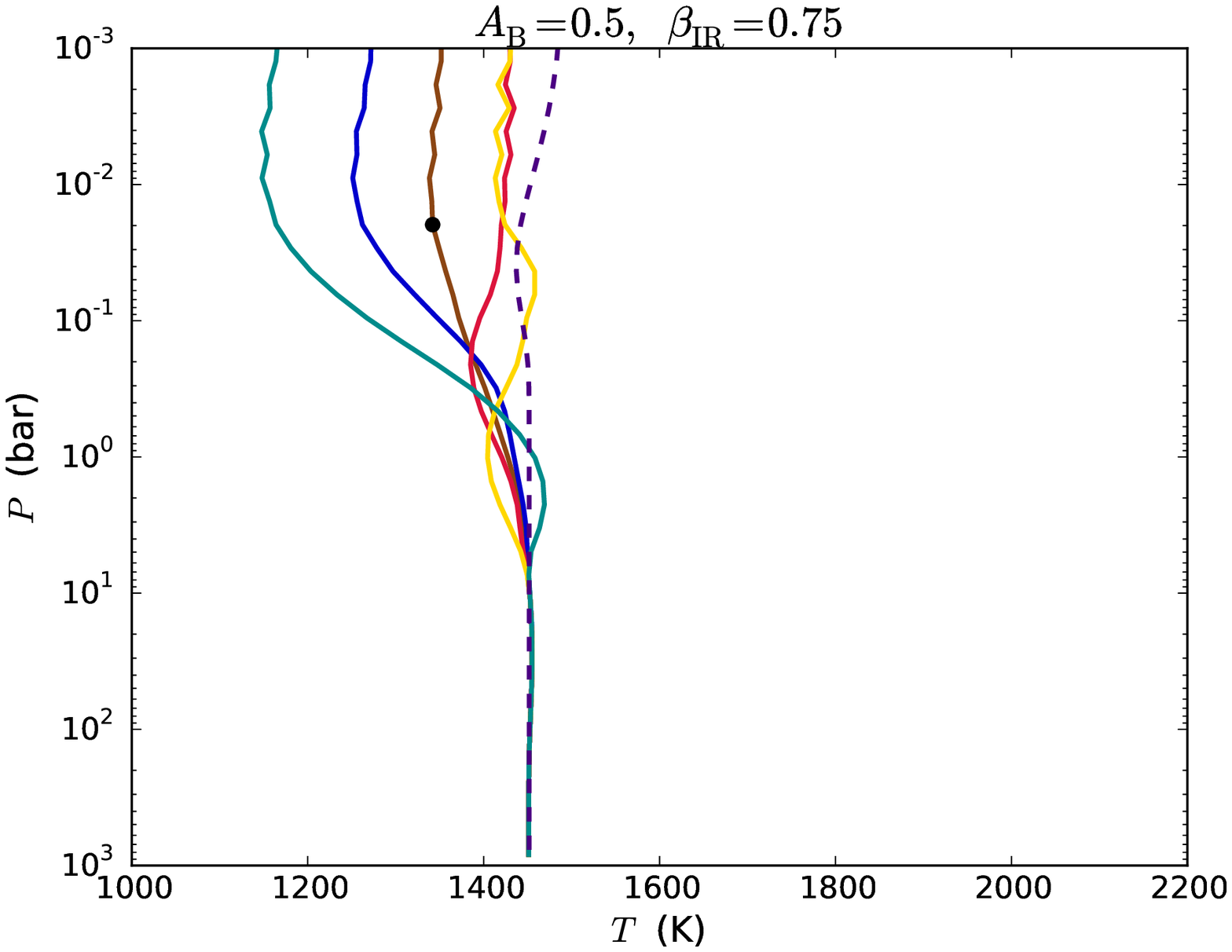}
\endminipage\hfill
\minipage{0.33\textwidth}
\includegraphics[width=\textwidth]{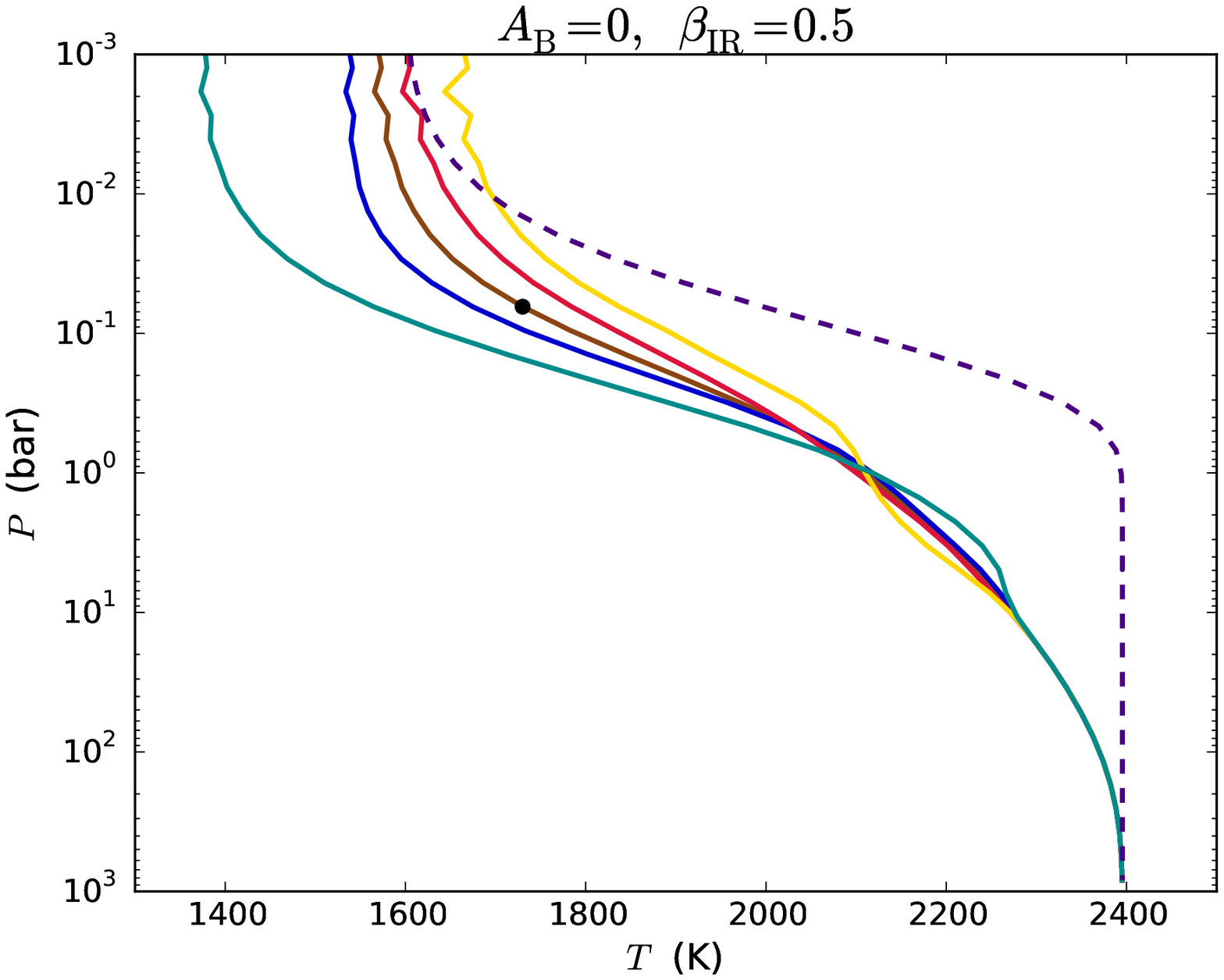}
\endminipage\hfill
\minipage{0.33\textwidth}
\includegraphics[width=\textwidth]{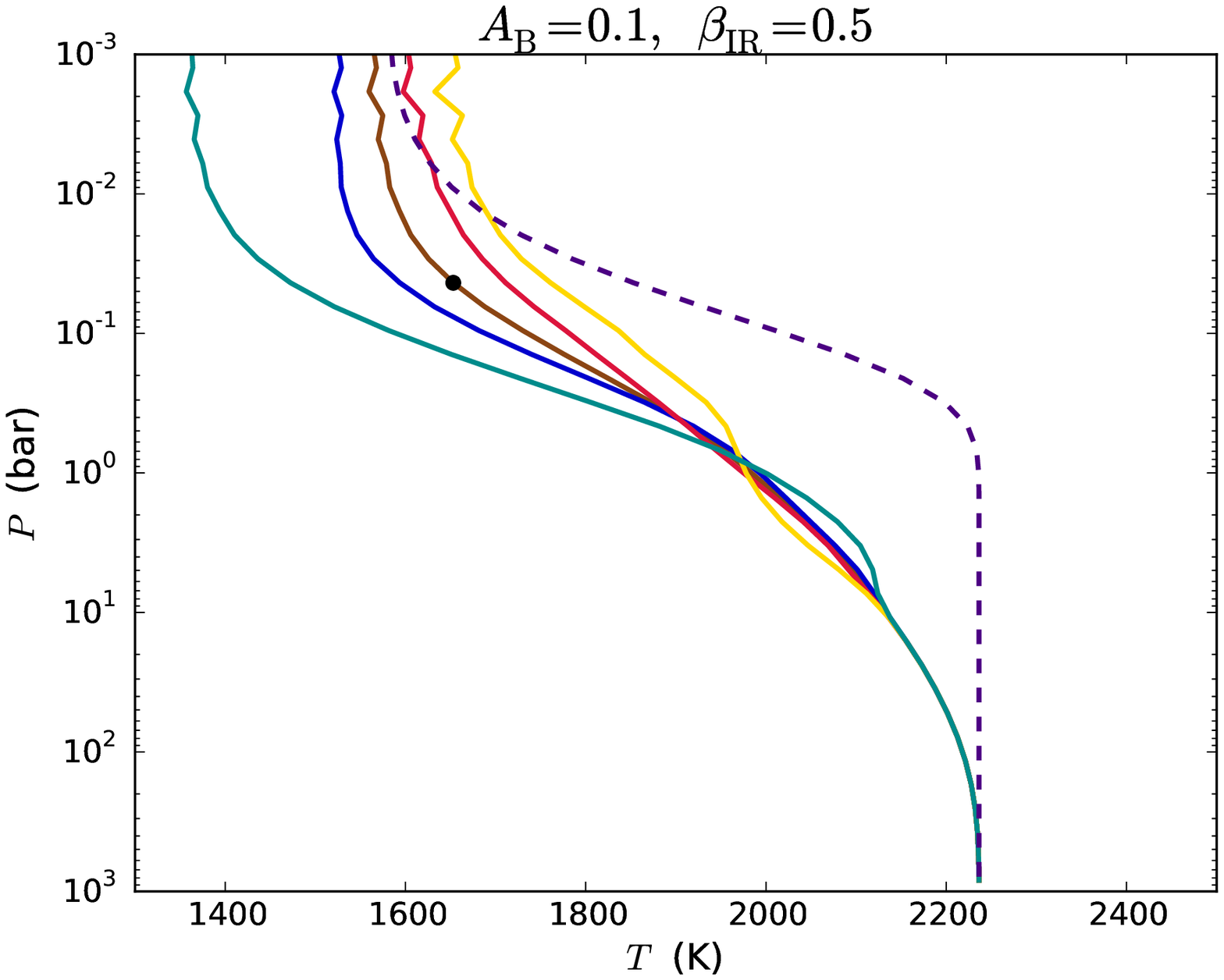}
\endminipage\hfill
\minipage{0.33\textwidth}
\includegraphics[width=\textwidth]{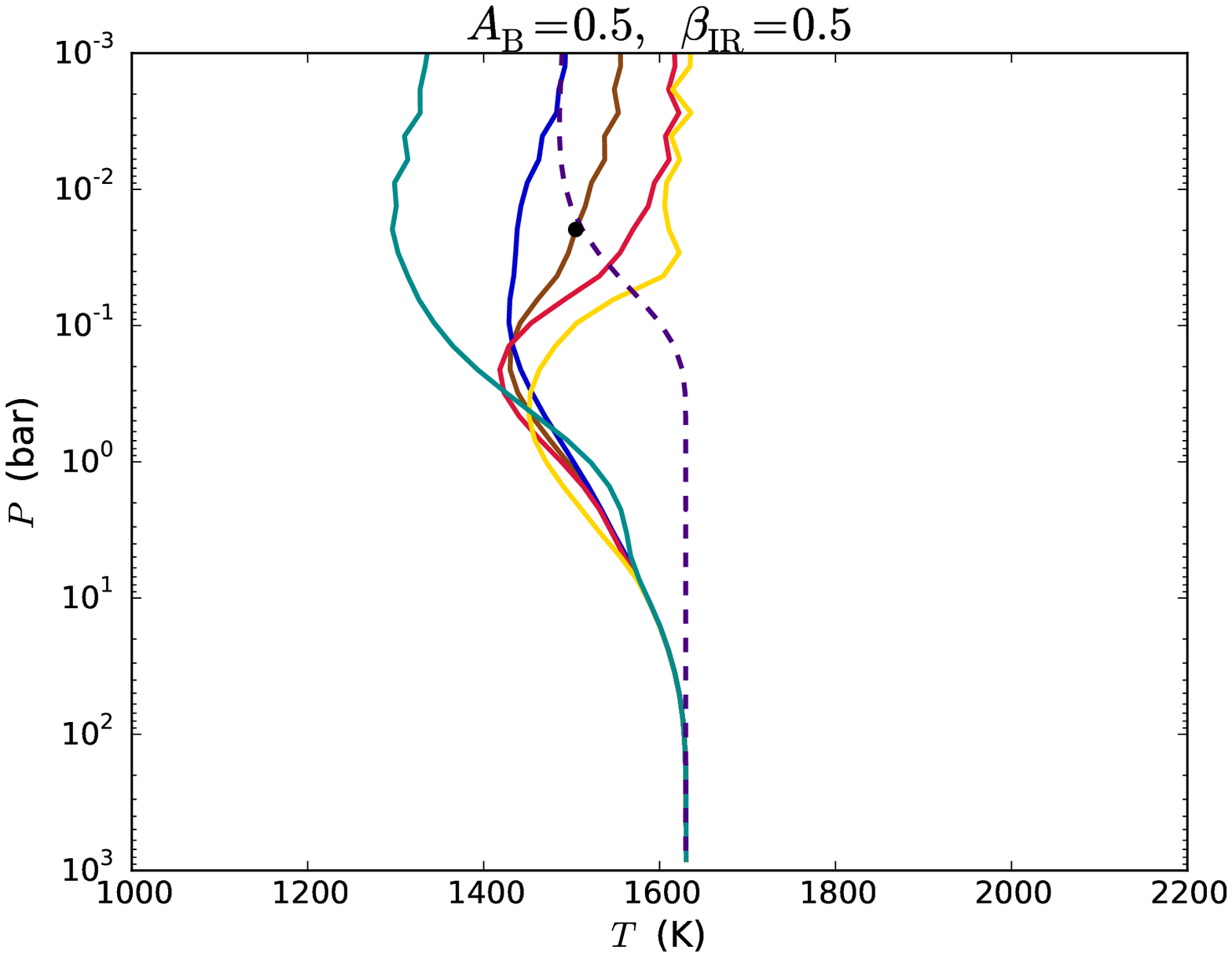}
\endminipage\hfill
\caption{Temperature-pressure profiles from our suite of GCMs.  The simulated profiles are shown as solid curves, while the dotted curve is taken from an analytical model (see text).  The various colours are for the global average (brown), nightside (blue), dayside (red), day-night teminator (yellow; 270$^\circ$ longitude) and night-day terminator (green; 90$^\circ$ longitude).  The black dots indicate the locations of the photon deposition depth or layer.}
\label{fig:tp}
\end{figure*}

The two-stream treatment of radiative transfer approximates the passage of radiation as a pair of incoming and outgoing rays (see \citealt{hml14} and references therein), which is a decent approximation if the physical thickness of the atmosphere is considerably smaller than the radius of the exoplanet.  Since we simulate 6 orders of magnitude in pressure, which corresponds to about 14 pressure scale heights and $H/R \sim 10^{-3}$--$10^{-2}$, even hot Jovian atmospheres are thin in terms of the simulation domain.  (Here, $H$ and $R$ denote the pressure scale height and the radius of the exoplanet, respectively.)

Computationally, one always needs to divide the model atmosphere into a discrete set of layers.  Upon being heated by starlight, each model layer has a finite temperature and emits a blackbody flux.  As the flux propagates out from the layer, it may be absorbed and re-emitted or scattered.  The blackbody fluxes from every layer need to be propagated throughout the atmosphere until the gradient of the net flux vanishes between the layers, such that the temperature reaches a steady state.  To execute this computational task requires that we have an analytical expression for propagating fluxes between a pair of layers, which may then be applied, pair-wise, to the entire atmosphere.  \cite{hml14} have previously derived analytical solutions for the incoming and outgoing fluxes for an arbitrary pair of layers,
\begin{equation}
\begin{split}
F_{\uparrow_i} =& \frac{1}{\left( \zeta_- {\cal T} \right)^2 - \zeta_+^2} \left\{ \left( \zeta_-^2 - \zeta_+^2 \right) {\cal T} F_{\uparrow_{i+1}} - \zeta_- \zeta_+ \left( 1 - {\cal T}^2 \right) F_{\downarrow_i} \right. \\
&+ \left. \pi B \left[ \zeta_- \zeta_+ \left( 1 - {\cal T}^2 \right) - \left( \zeta_-^2 {\cal T} + \zeta_+^2 \right) \left( 1 - {\cal T} \right) \right] \right\},  \\
F_{\downarrow_{i+1}} =& \frac{1}{\left( \zeta_- {\cal T} \right)^2 - \zeta_+^2} \left\{ \left( \zeta_-^2 - \zeta_+^2 \right) {\cal T} F_{\downarrow_i} - \zeta_- \zeta_+ \left( 1 - {\cal T}^2 \right) F_{\uparrow_{i+1}} \right. \\
&+ \left. \pi B \left[ \zeta_- \zeta_+ \left( 1 - {\cal T}^2 \right) - \left( \zeta_-^2 {\cal T} + \zeta_+^2 \right) \left( 1 - {\cal T} \right) \right] \right\},  \\
\end{split}
\label{eq:twostream_solutions}
\end{equation}
where the index $i$ refers to the $i$-th layer in our model atmosphere.  (Our convention is that higher values of $i$ correspond to higher pressures.)  These solutions generalise the pure-absorption ones previously implemented by \cite{frierson06} in the FMS, which formed the computational basis for \cite{hfp11} and \cite{php12}.  The Planck or blackbody function is denoted by $B$.  As in the two-stream approximation, equation (\ref{eq:twostream_solutions}) describes the radial transfer of radiation only.

In equation (\ref{eq:twostream_solutions}), there are several dimensionless quantities that need to be specified for each layer.  First, the transmission function (${\cal T}$) quantifies the transparency or opaqueness of each layer to radiation,
\begin{equation}
{\cal T} = \exp{\left( - \alpha ~\Delta \tau \right)},
\end{equation} 
where $\Delta \tau = \tau_{i+1} - \tau_i$ is the difference in optical depth between a pair of layers.  The coefficient in the exponent is \citep{hml14}
\begin{equation}
\alpha = 2\beta_{\rm IR}\left( \frac{1-g_0}{1-g_0 \beta_{\rm IR}^2} \right),
\label{eq:alpha}
\end{equation}
where $\beta_{\rm IR}$ is the scattering parameter in the longwave/infrared.  It is related to the single-scattering albedo ($\omega_0$) and scattering asymmetry factor ($g_0$) via \citep{hml14}
\begin{equation}
\beta_{\rm IR} = \left( \frac{1-\omega_0}{1-\omega_0 g_0} \right)^{1/2}.
\end{equation} 
In a departure from \cite{frierson06}, the factor of 2 in equation (\ref{eq:alpha}) originates from demanding that an opaque, isothermal atmosphere produces $\pi B$ of flux in each hemisphere.  The factor of 2 is sometimes termed the ``diffusivity factor" and it is also the reciprocal of the first Eddington coefficient \citep{hml14}.

Second, the (dimensionless) coupling coefficients are \citep{hml14}
\begin{equation}
\zeta_\pm = \frac{1 \pm \beta_{\rm IR}}{2},
\end{equation}
so named because they allow the boundary conditions incident upon each layer to be coupled in the presence of scattering.  In the limit of pure absorption, we have $\zeta_+=1$ and $\zeta_-=0$ and the solutions in equation (\ref{eq:twostream_solutions}) decouple in the sense that they may be solved independently of each other.

In the present study, we will only consider small aerosols, which have particle radii that are smaller than the wavelength of infrared emission.  This allows us to set $g_0=0$ (isotropic scattering in the infrared) and thus obtain $\alpha = 2 \beta_{\rm IR}$ and $\beta_{\rm IR} = (1 - \omega_0 )^{1/2}$.  We set the radius of our spherical, monodisperse aerosols to be $r=1$ $\mu$m.  The Stokes numbers associated with our micron-sized aerosols are typically much larger than unity, which justifies our approximation of using a single-fluid GCM to study them.

\subsection{Implementation within FMS and benchmarking}

We implement equations (\ref{eq:beers}) and (\ref{eq:twostream_solutions}) into the radiative transfer module of the FMS.  In the visible, the implementation of the generalised Beer's law involves specifying the stellar flux incident upon the top of the model atmosphere ($F_{\rm TOA}$) and computing its diluted values, across the radial grid, according to equation (\ref{eq:beers}).  This procedure is repeated for each radial column of the model atmosphere.

For infrared radiation, we use equation (\ref{eq:twostream_solutions}) to perform radiative transfer for each radial column of atmosphere.  In the pure absorption limit, the arrays for the incoming ($F_{\downarrow_{i+1}}$) and outgoing ($F_{\uparrow_i}$) fluxes may be computed independently of each other and depend only on the boundary conditions at the top ($F_{\downarrow_i}$) and bottom ($F_{\uparrow_{i+1}}$), respectively, of each atmospheric layer.  For example, the outgoing flux may be computed by propagating the bottom boundary condition upwards using the second equation in (\ref{eq:twostream_solutions}), without knowledge of the incoming flux.  Similarly, the array for the incoming flux may be computed.  When scattering is present, each flux array now depends on both boundary conditions and thus cannot be computed independently of the other.  This forces us to adopt an iterative approach.  For example, for the outgoing flux we begin at the bottom of the simulation domain and compute $F_{\uparrow_i}$ for all $i$.  Initially, we set $F_{\downarrow_i}=0$.  Upon populating the $F_{\uparrow_{i+1}}$ array, we use it to compute the $F_{\downarrow_{i+1}}$ array.  We then use $F_{\downarrow_{i+1}}$ to update the $F_{\uparrow_i}$ array.  We iterate until the fractional difference in the fluxes is less than $10^{-4}$, which typically requires about 10 iterations.

To test that we are implementing the FMS correctly, we reproduce the pure-absorption GCMs of \cite{php12} and verify that we are able to reproduce their climatology plots (zonal-mean zonal wind, temperature, potential temperature and Eulerian-mean streamfunction).  (We do not show the reproduced figures in the present paper.)  We use the same values of hyperviscosity as for Model H in \cite{php12}.

\subsection{Optical phase curves, condensation curves and infrared phase curves}

\begin{figure*}
\includegraphics[width=\columnwidth]{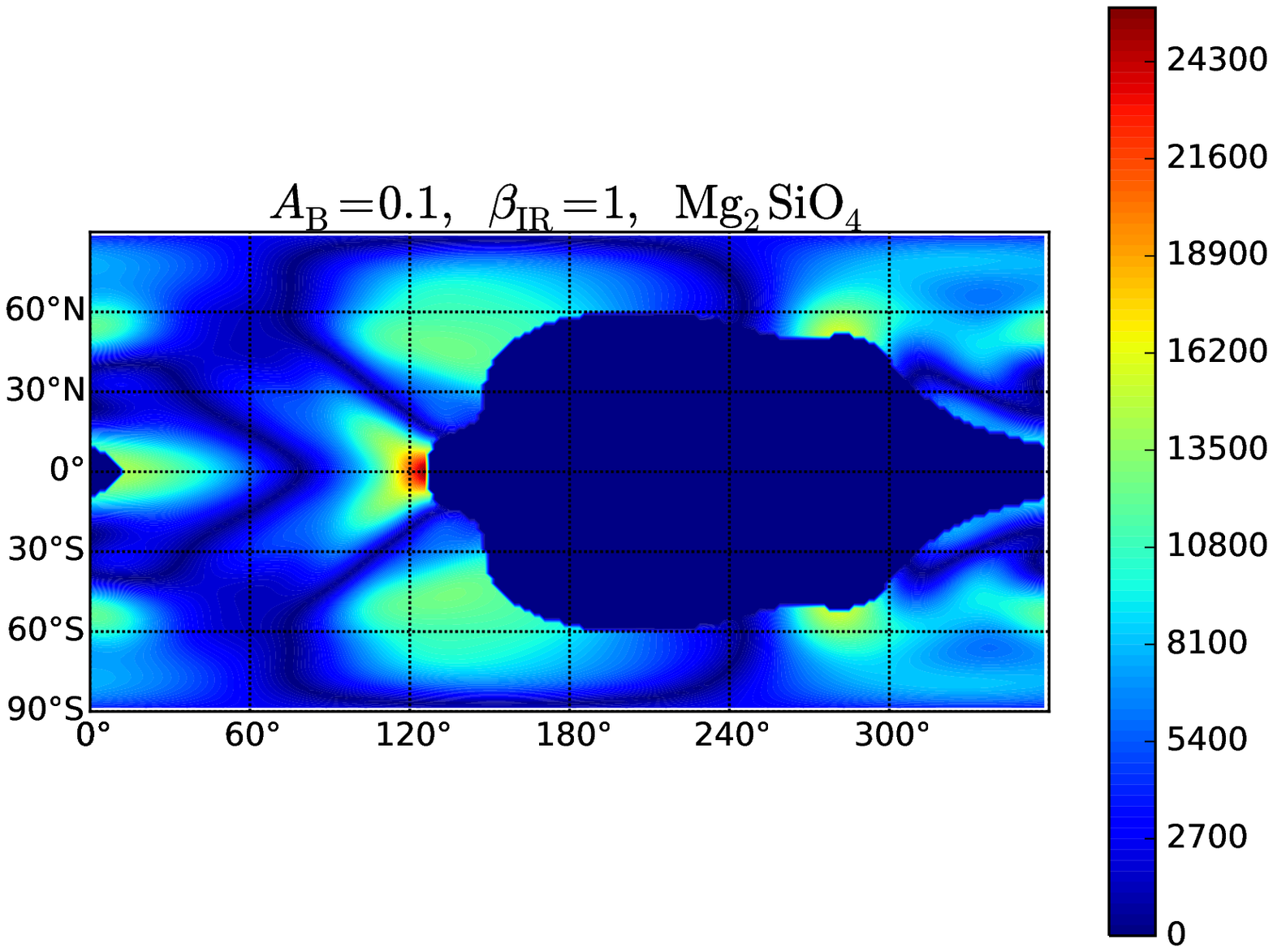}
\includegraphics[width=\columnwidth]{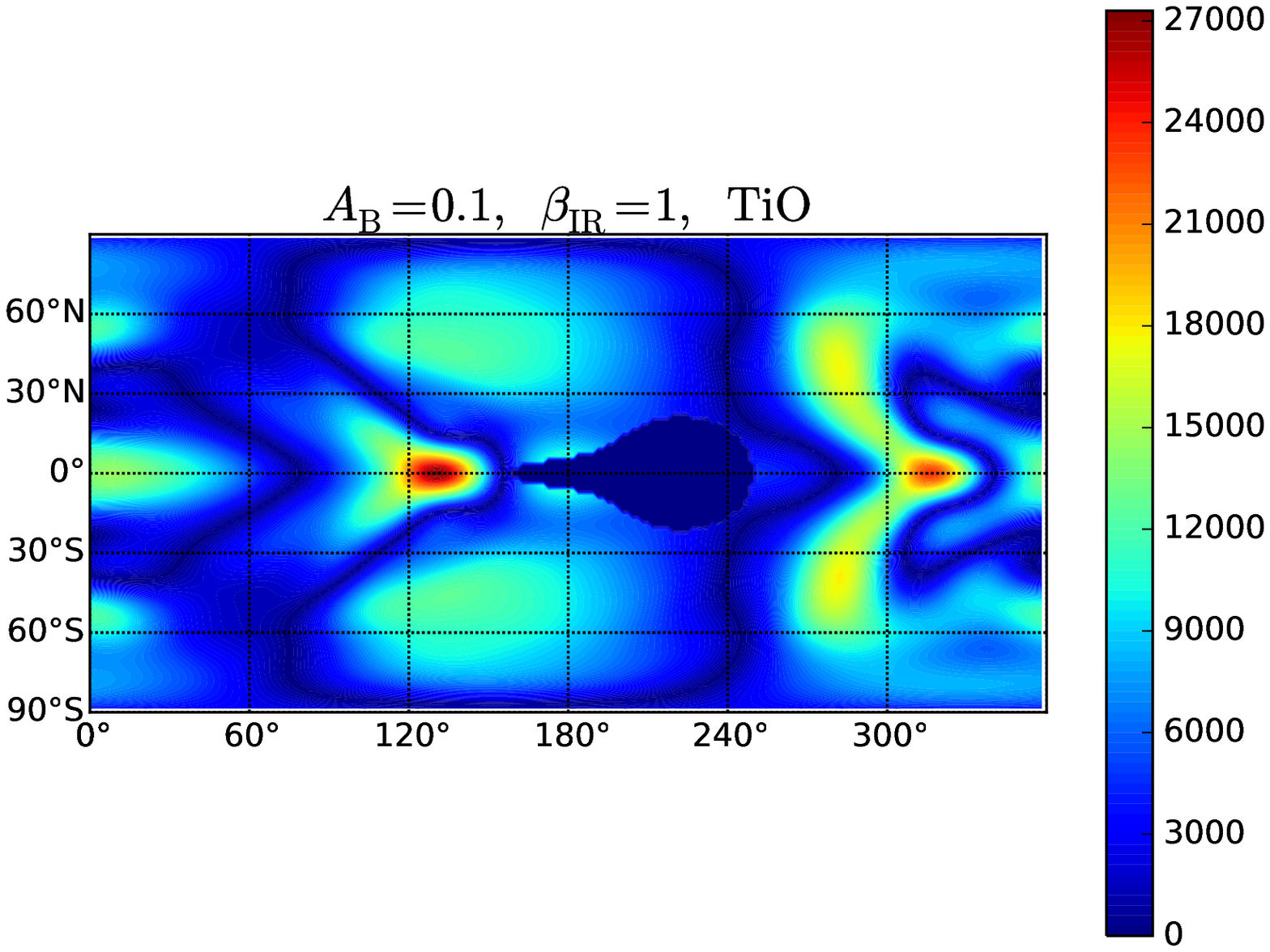}
\vspace{-0.1in}
\caption{Maps of the Stokes number as a function of latitude and longitude.  Regions with temperatures that are too high for condensation are excluded (see text).  To illustrate the dependence on composition, we compare the Stokes maps associated with forsterite (left panel) and titanium oxide (right panel).  The substellar point is located at a longitude of 180$^\circ$.}
\vspace{-0.1in}
\label{fig:stokes}
\end{figure*}

For each GCM, we postprocess its output to compute the optical phase curves in an approximate way.  We accomplish this by calculating the Stokes number,
\begin{equation}
S = \frac{v_z}{v_{\rm term}},
\end{equation}
where $v_z$ is the vertical/radial component of the velocity and the terminal velocity is given by (see \citealt{spiegel09} and references therein)
\begin{equation}
v_{\rm term} =\frac{2 {\cal C} r^2 \rho_{\rm int} g}{9 \rho \nu},
\end{equation}
where ${\cal C}$ is a correction factor that depends on the Knudsen number, $\rho_{\rm int}$ is the internal mass density of the aerosols, $\rho$ is the mass density of the atmospheric gas and $\nu$ is the kinetic viscosity.  (See \citealt{hd13} on how to compute ${\cal C}$ and $\nu$.)  We set $\rho_{\rm int} = 3$ g cm$^{-3}$, although this is of no consequence to our computed optical phase curves as we normalise $S$ to have a maximum value of unity.  We do not attempt to model the exact abundance of the aerosols or condensates and instead use $S$ as a proxy for their relative abundance, which is in turn related to the local reflectivity of the atmosphere.  By normalising $S$ to have a maximum value of unity, we may use it to study the \textit{shape} of the optical phase curve as the free parameters are varied.

We evaluate $S$ at the photon deposition depth, which is the atmospheric layer where starlight is mostly absorbed and is located at a pressure of \citep{hhps12,hml14}
\begin{equation}
P_{\rm D} = \frac{0.63g}{\kappa_{\rm S_0}}\left( \frac{1-A_{\rm B}}{1+A_{\rm B}}\right).
\end{equation}
We have $P_{\rm D} \approx 55, 45$ and 18 mbar for $A_{\rm B}$=0, 0.1 and 0.5, respectively.

At the photon deposition depth, we exclude regions where the temperature is too high for particles to condense out, which depends on the assumed composition of the aerosol or condensate.  We use the condensation curves of \cite{burrows06} and consider the following compositions: corundum (Al$_2$O$_3$), enstatite (MgSiO$_3$), forsterite (Mg$_2$SiO$_4$), iron (Fe) and titanium oxide (TiO).  Figure \ref{fig:condense} displays the condensation curves used in the present study.

For both the optical and infrared phase curves, we use the method of \cite{ca08} to transform flux or $S$ maps (which are functions of latitude and longitude) into phase curves (which are functions of longitude).  The same approach was used in \cite{hfp11} for calculating infrared phase curves.  The main difference is that, for the optical phase curves, we ignore the contributions of the $S$ maps from the nightside.

\section{Results}
\label{sect:results}

\begin{figure*}
\includegraphics[width=\columnwidth]{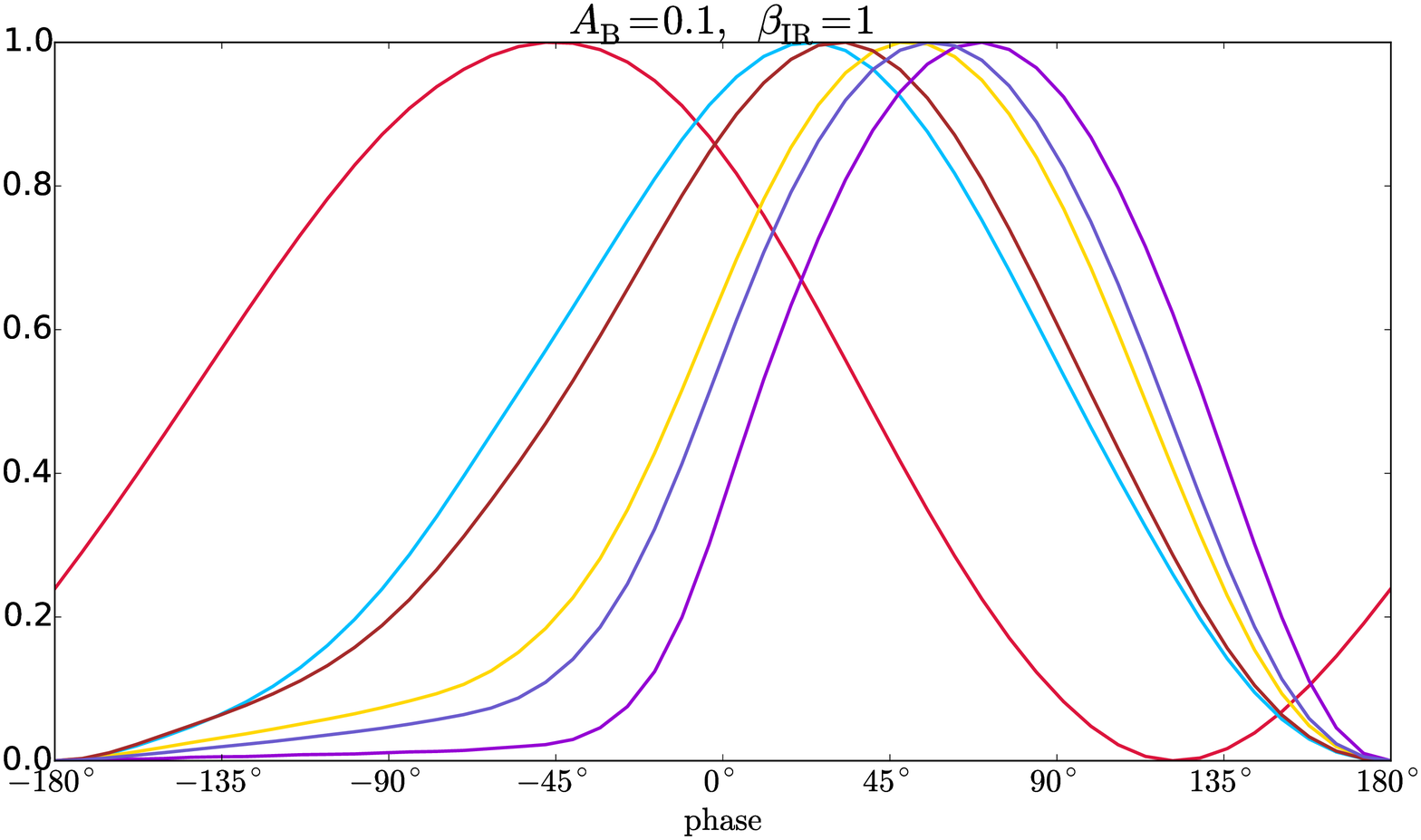}
\includegraphics[width=\columnwidth]{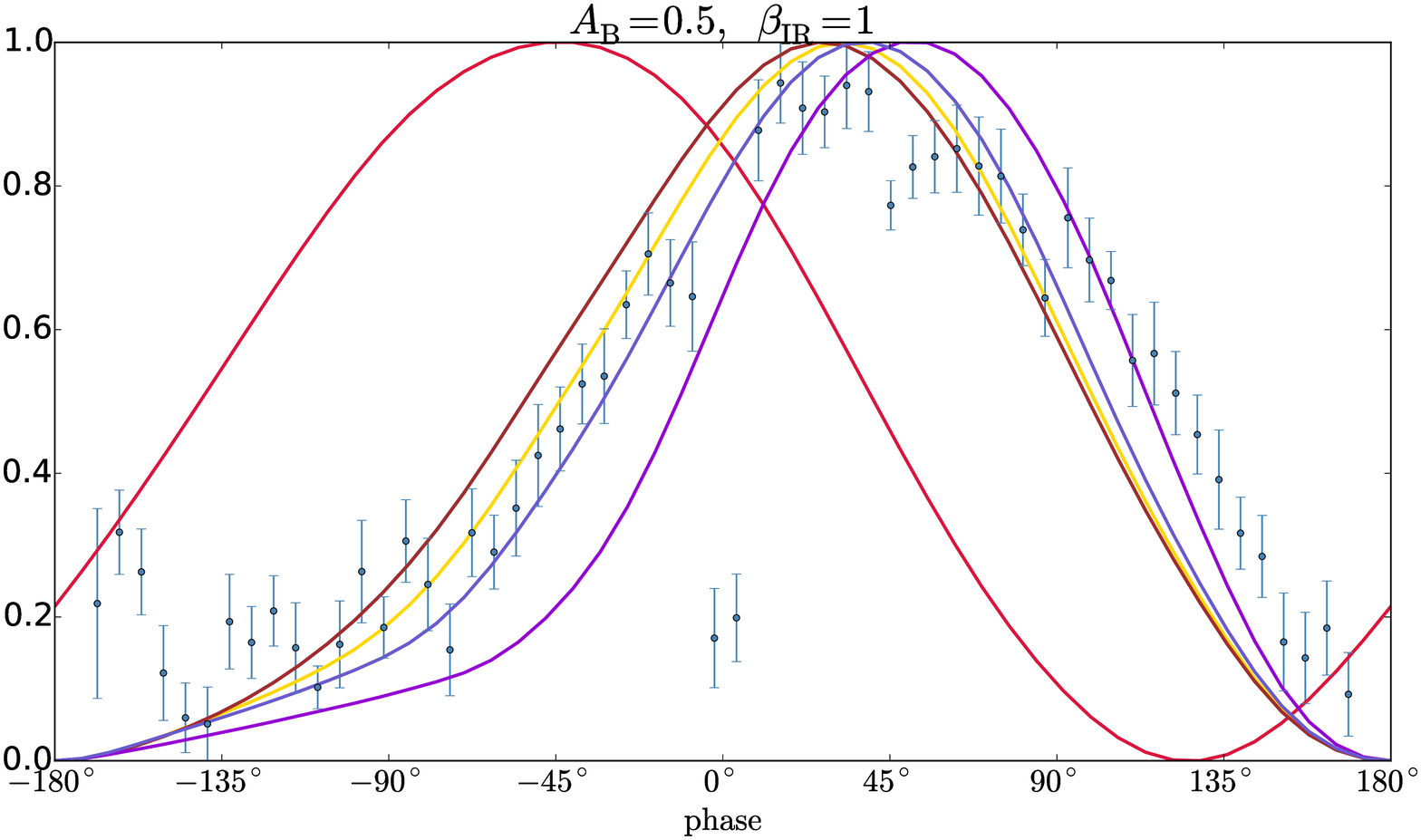}
\includegraphics[width=\columnwidth]{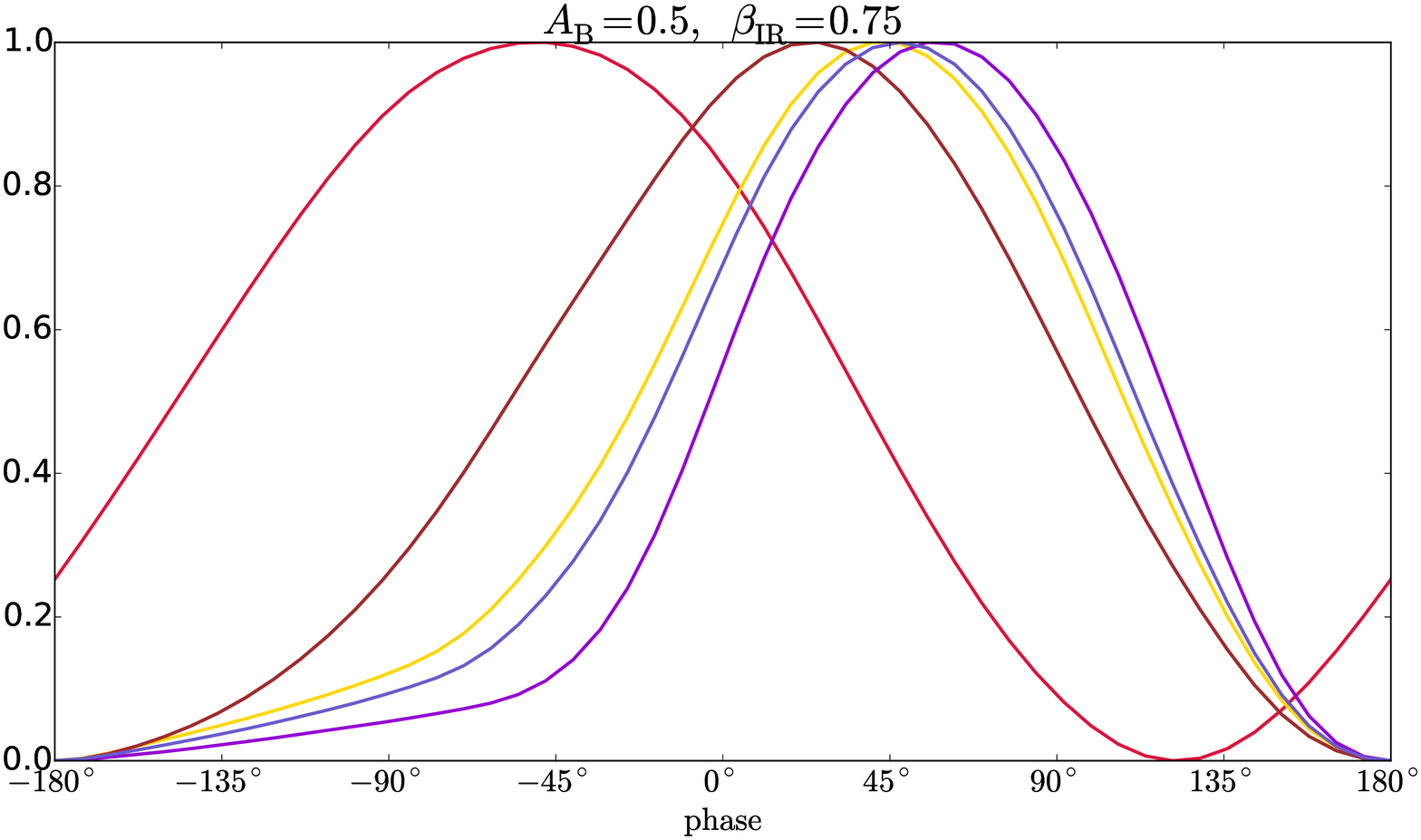}
\includegraphics[width=\columnwidth]{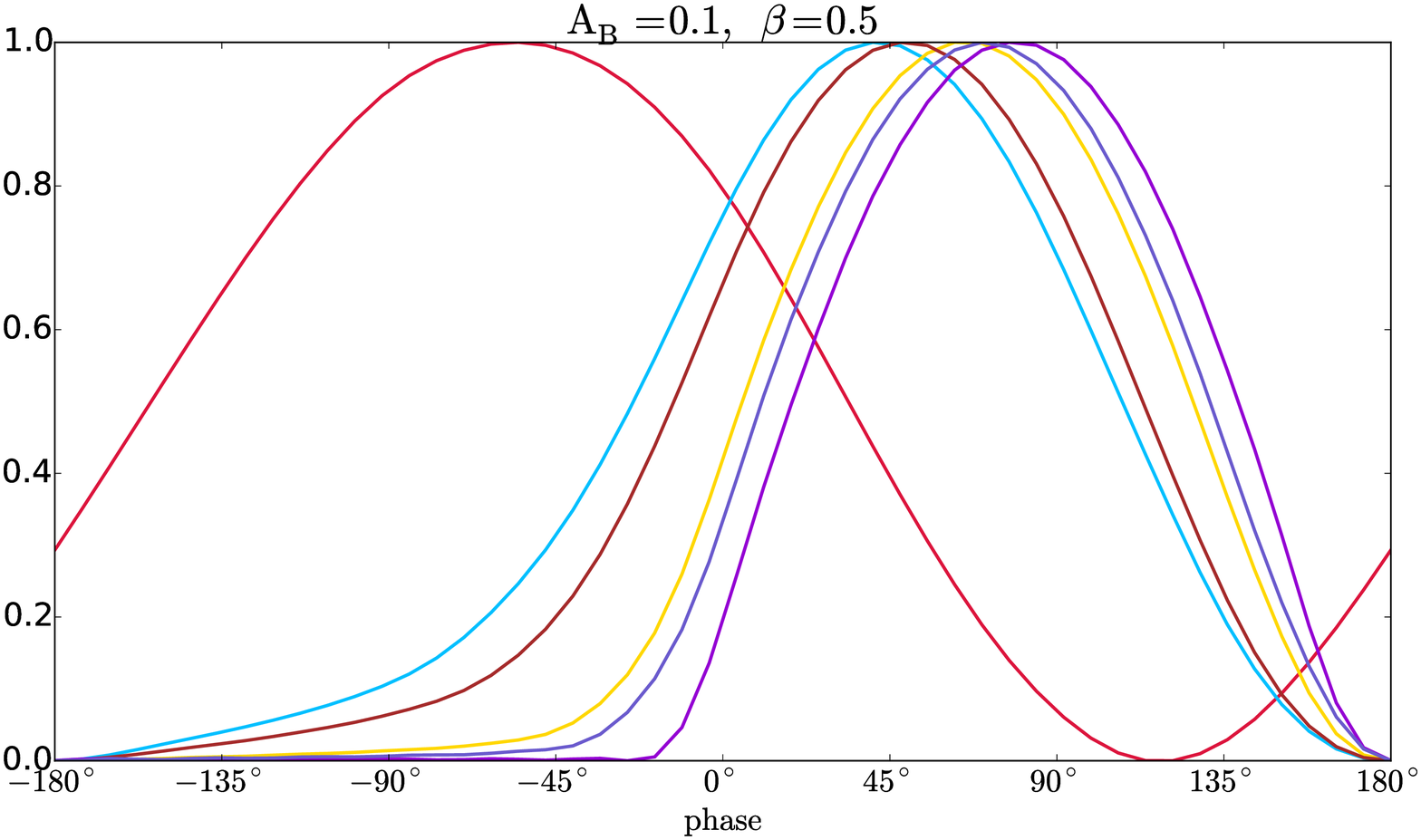}
\caption{Infrared and optical phase curves from four of our GCMs, normalised to unity to focus on the shape of the curves.  The measured optical phase curve has been normalised too.  Within each panel, we show the optical phase curves associated with various aerosol species.  The red curve that has a negative peak offset is the infrared phase curve.  The other curves are the optical phase curves; the assumed composition of the aerosol species is colour-coded in the same way as for Figure \ref{fig:phase_offset}.  For the panels with $A_{\rm B}=0.5$, the optical phase curves for corundum and titanium oxide coincide, because of the low atmospheric temperatures.  A direct comparison to data should only be made for $A_{\rm B}=0.5$ and $\beta_{\rm IR}=1$ (see text).}
\label{fig:phase_curves}
\end{figure*}

\begin{figure*}
\includegraphics[width=1.8\columnwidth]{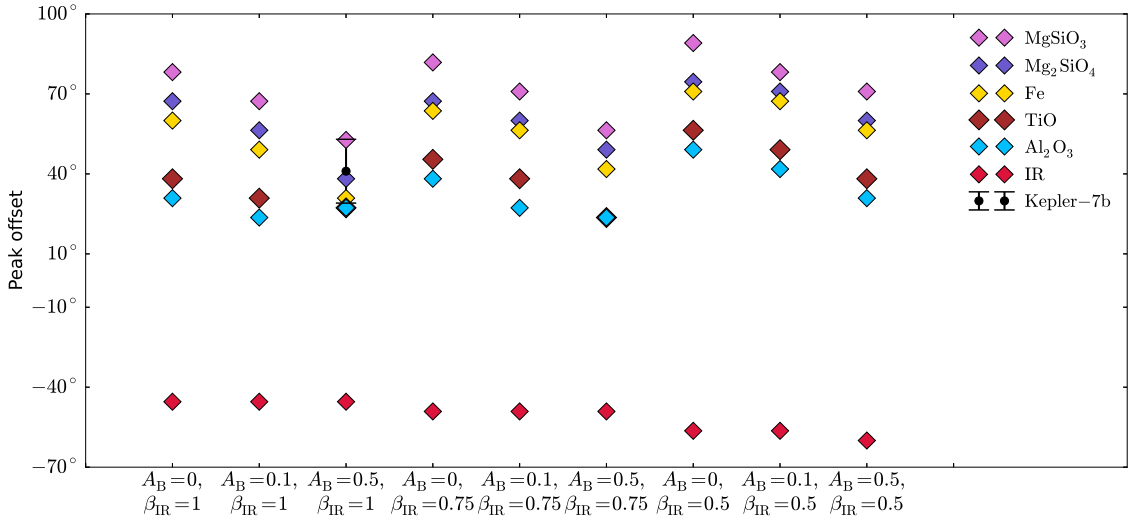}
\caption{Peak offsets of the optical (marked by the various aerosol species) and infrared (marked by ``IR") phase curves for our suite of GCMs.  Again, a direct comparison to data should only be made for $A_{\rm B}=0.5$ and $\beta_{\rm IR}=1$ (see text).}
\label{fig:phase_offset}
\end{figure*}

We now present a suite of simulations tailored to the hot Jupiter Kepler-7b.  The equilibrium temperature is $T_{\rm eq}=1630$ K \citep{esteves15}, which translates into a stellar constant of $F_{\rm TOA} =1.6\times10^9$ erg cm$^{-2}$ s$^{-1}$.  The surface gravity of Kepler-7b is $g=437$ cm s$^{-2}$ and its white-light radius is $R=1.622 ~R_{\rm J} = 1.1596 \times 10^{10}$ cm, where $R_{\rm J} = 7.1492 \times 10^9$ cm is the equatorial radius of Jupiter \citep{esteves15}.  We assume a constant infrared opacity of $\kappa_{\rm IR} = 0.01$ cm$^2$ g$^{-1}$, which translates into an infrared photospheric pressure of about 44 mbar.  We assume Kepler-7b to be tidally locked and take its rotational frequency to be equal to its orbital frequency: $\Omega = 1.5 \times10^{-5}$ s$^{-1}$ \citep{esteves15}.  

Overall, each GCM has only four free parameters: the shortwave opacity normalisation ($\kappa_{\rm S_0}$), the infrared opacity, the Bond albedo and the infrared scattering parameter.  The stellar constant and surface gravity are not considered to be free parameters, because their values are fixed by the observations of Kepler-7b.  For each simulation, we adopt an initial, constant temperature, computed using \citep{hml14}
\begin{equation}
T_{\rm init} = \left[ \frac{F_{\rm TOA} \left( 1 - A_{\rm B} \right)}{8\sigma_{\rm SB}}\left( \frac{4}{3}+\frac{\kappa_{\rm IR}\beta_{\rm S}}{\kappa_{\rm S_0} \beta_{\rm IR}^2}\right) \right]^{1/4},
\end{equation}
where $\sigma_{\rm SB}$ is the Stefan-Boltzmann constant, and list its values for our suite of models in Table \ref{tab:params}.  In the absence of better knowledge, we start each simulation from a state of rest.

\begin{table}
\centering
\caption{Initial temperatures used for suite of Kepler-7b simulations}
\label{tab:params}
\begin{tabular}{lcc}
\hline
$A_{\rm B}$ & $\beta_{\rm IR}$ & $T_{\rm init}$ (K) \\
\hline
 0 & 1& 1852\\
 0.1 & 1& 1752\\
 0.5 & 1& 1370\\
 0 & 0.75& 2038\\
 0.1 & 0.75& 1916\\
 0.5 & 0.75& 1451\\
 0 & 0.5 & 2395\\
 0.1& 0.5& 2236\\
 0.5& 0.5& 1630\\
\hline
\end{tabular}\\
Note that $\beta_{\rm IR} = 1$, $0.75$ and $0.5$ correspond to $\omega_0 = 0, \approx 0.44$ and $= 0.75$, respectively.
\end{table}

\subsection{Climatology of GCM suite}

Figures \ref{fig:zonalwind}, \ref{fig:temperature} and \ref{fig:streamfunction} show the climatology of our suite of GCMs.  Across variations in the optical and infrared scattering, the model atmosphere maintains an equatorial zonal jet that penetrates to $\sim 1$ bar (Figure \ref{fig:zonalwind}).  In agreement with \cite{hfp11}, the equatorial zonal jet exists only where the potential temperature varies across latitude.  The dynamically inert part of the atmosphere ($\gtrsim 1$ bar), where no zonal jet exists, is characterised by constant potential temperature (and hence entropy) across latitude.  We next construct the Eulerian-mean streamfunction by integrating the meridional component of the velocity down to $\sim 1$ bar, which reveals equator-to-pole circulation cells that are robust to variations in the scattering of starlight and thermal emission (Figure \ref{fig:streamfunction}).  

The temperature across pressure and latitude changes as $A_{\rm B}$ and $\beta_{\rm IR}$ are varied, according to trends predicted by one-dimensional analytical models (\citealt{hhps12,hml14}; Figure \ref{fig:temperature}).  Generally, increasing $A_{\rm B}$ and $\beta_{\rm IR}$ cools and warms the atmosphere, respectively.  Figure \ref{fig:tp} compares the various one-dimensional temperature-pressure profiles, across pressure, with analytical, globally-averaged models from \cite{hml14}.  Since both the simulated and analytical profiles derive from the same governing equations, any discrepancies may be attributed to localised differences in stellar heating and/or atmospheric dynamics.

\subsection{Optical and infrared phase curves}

Our calculation of the optical phase curves starts from the evaluation of the Stokes number at the photon deposition depth.  In Figure \ref{fig:stokes}, we illustrate this procedure and how the detailed structure of $S$ depends on the assumed aerosol composition.  The infrared photosphere of an atmosphere is a direct probe of the temperature across latitude and longitude.  By contrast, the reflectivity of the atmosphere at the photon deposition depth depends on condensation physics (via the condensation curves, which determine if a specific species of aerosol will condense out of the gas) and atmospheric dynamics (via the vertical/radial component of the velocity, which determines if an aerosol particle of a given size may be lofted).  Thus, we expect infrared and optical phase curves to be complementary probes of an exoplanetary atmosphere.

Figure \ref{fig:phase_curves} compares the infrared and optical phase curves for four of our GCMs (with non-zero albedos).  The photometric data shown are the original Kepler-7b phase-curve photometry from \cite{demory13} that have been binned per 2 hours for clarity and normalised.  We note that Kepler-7b has $A_{\rm B} \approx 0.5$ (assuming isotropic scattering by aerosols or condensates; \citealt{demory11,demory13,hd13}).  Furthermore, our assumption of micron-sized aerosols implies $\beta_{\rm IR} \approx 1$.  Nevertheless, we show GCMs with other values of $A_{\rm B}$ and $\beta_{\rm IR}$ so that these parameter dependences may be elucidated.  When confronted by optical phase curve data of Kepler-7b from \cite{demory13}, we see that our simple treatment for obtaining optical phase curves succeed in reproducing the measured peak offset.  It roughly produces the correct overall shape of the optical phase curves, but our computed curves have widths that are somewhat too narrow if a single aerosol/condensate composition is assumed.  Notwithstanding, the peak offset of the optical phase curves is sensitive to the assumed composition of the aerosols, as demonstrated in Figures \ref{fig:phase_curves} and \ref{fig:phase_offset}.  This property may be considered tentative until it is checked by more sophisticated calculations that include multiple scattering.  By contrast, the peak offset of the infrared phase curves is somewhat insensitive to variations in both optical and infrared scattering (Figure \ref{fig:phase_offset}) and therefore yields little information on the aerosol properties.

A robust outcome of the GCMs is that infrared phase curves always peak eastwards, as has been shown in previous studies \citep{sg02,showman09,sp11,hfp11,tsai14}, but optical phase curves peak westwards because they probe regions of the atmosphere that are cool enough to form aerosols or condensates.  We predict that the peak offset of the infrared phase curve of Kepler-7b is 45$^\circ$ if the aerosols/condensates are small ($\beta_{\rm IR}=1$).  To within the measurement uncertainties, the measured peak offset ($41^\circ \pm 12^\circ$; \citealt{demory13}) of the optical phase curve of Kepler-7b is consistent with all of the aerosol species considered in the current study (corundum, enstatite, forsterite, iron, titanium oxide; Figure \ref{fig:phase_offset}).  This motivates the need for more precise measurements of optical phase curves.  We note that \cite{webber15} used a planetary albedo model, coupled to a one-dimensional, plane-parallel radiative transfer model without atmospheric dynamics, to conclude that iron clouds are too dark to fit the observed magnitudes of the phase curve.  Given the limitation of our modeling method, we cannot directly compute and predict the phase curve magnitude and thus can neither corroborate nor refute this conclusion.

Generally, we conclude that optical phase curves offer complementary and potentially decisive constraints on the composition of aerosols or condensates in cloudy exoplanetary atmospheres.

\section{Discussion}
\label{sect:discussion}

\subsection{Caveats, comparison with previous work and opportunities for future work}

In the present study, we have implemented a simple but plausible treatment of both optical and infrared scattering in three-dimensional GCMs of the hot Jupiter Kepler-7b and used them to study the influence of aerosol composition on the shape of the optical and infrared phase curves.  Our study is complementary to that of \cite{par13}, who did not include the radiative effects of the aerosols on the atmosphere, but performed a more realistic treatment of their dynamical coupling with the atmospheric flow, albeit in the pure-absorption limit.  Clearly, the way forward is to construct GCMs that deal with the radiative effects of the aerosols and their dynamical coupling to the atmosphere in a more realistic way.  

Eventually, one would need to include realistic models of how the aerosols themselves would form out of the atmospheric gas via a detailed treatment of the chemistry.  The chemistry would have to be modelled self-consistently with the molecular and aerosol opacities to determine the global temperature structure of the atmosphere.  Such a model would self-consistently predict the abundance of the aerosols relative to the gas, their distribution of sizes and also their spatial distribution throughout the atmosphere without having to parametrise these quantities.  \cite{lee15} made initial strides in this direction by using three-dimensional GCMs, albeit executed on a truncated (non-global) grid, as background states for post-processing cloud formation calculations, but did not explicitly model optical phase curves.  Like in the present study, they employed two-stream radiative transfer and used prescribed gas opacities, implying that the gas and cloud chemistry and opacities are not modelled self-consistently.  They also did not model horizontal mixing.  Clearly, there are ample opportunities for future work.

\subsection{The need for both infrared and optical phase curves for each exoplanet}

To date, we do not have measured infrared \textit{and} optical phase curves for the same exoplanet.  Such a measurement was attempted for Kepler-7b, but even the secondary eclipses were undetected in the infrared using the Spitzer Space Telescope \citep{demory13}.  Our present study has demonstrated that infrared and optical phase curves offer complementary information on the atmosphere of the exoplanet.  Specifically, optical phase curves encode information on the composition of the aerosols or condensates contained within a cloudy atmosphere, which will reduce the degeneracy associated with interpreting its infrared spectra \citep{lee14,ws15}.  However, our study has also motivated the need for measuring optical phase curves to even higher precision.  Future telescopes, such as CHEOPS (Characterising Exoplanets Satellite) of the European Space Agency, will offer opportunities for recording high-precision optical phase curves that will complement infrared spectra and phase curves from the James Webb Space Telescope.

\section*{Acknowledgments}

MO and KH thank the PlanetS NCCR framework, the Swiss National Science Foundation and the Swiss-based MERAC Foundation for financial support.  MO thanks Hans Martin Schmid for supporting the project from the side of Institute for Astronomy at ETH Z\"{u}rich.  We are grateful to Olivier Byrde and his crack team for maintaining the Brutus computing cluster of ETH Z\"{u}rich and for providing prompt and competent technical support.  We thank Jo\~{a}o Mendon\c{c}a, Luc Grosheintz, Matej Malik and Daniel Kitzmann for technical assistance and advice.

% Don't change these lines
\bsp	% typesetting comment
\label{lastpage}
\end{document}